\documentclass[12pt]{article}
 \input{epsf.sty}
 \input{epsf.tex}
 \newcommand{\be}{\begin{equation}}
 \newcommand{\ee}{\end{equation}}
 \newcommand{\ba}{\begin{eqnarray}}
 \newcommand{\ea}{\end{eqnarray}}
 \def\Journal#1#2#3#4{{#1} {#2}, #3 (#4)}

 \def\ASP{Astropart.\ Physics}
 \def\CQG{Class.\ Quantum Grav.}
 \def\GRG{Gen.\ Rel. Grav.}
 \def\IJMPA{Int.\ J.\ Mod.\ Phys.\ A}
 \def\IJMPD{Int.\ J.\ Mod.\ Phys.\ D}
 \def\JMP{J.\ Math.\ Phys.}
 \def\MPLA{Mod.\ Phys.\ Lett.\ A}
 \def\NPB{Nucl.\ Phys.\ B}
 
 \def\PLB{Phys.\ Lett.\  B}
 \def\PRL{Phys.\ Rev.\ Lett.}
 \def\PRD{Phys.\ Rev.\ D}
 \def\PRP{Phys.\ Rep.}
 
 \def\laq{\,\raise 0.4ex\hbox{$<$}\kern -0.8em\lower 0.62ex\hbox{$\sim$}\,}
 \def\gaq{\,\raise 0.4ex\hbox{$>$}\kern -0.7em\lower 0.62ex\hbox{$\sim$}\,}
 \def\Tb{\tau_{\scriptscriptstyle BCLN}}
 \def\xb{x_{\scriptscriptstyle BCLN}} 
 \def\yb{y_{\scriptscriptstyle BCLN}}
 \def\zb{z_{\scriptscriptstyle BCLN}}
 \def\Ob{\Omega_{\scriptscriptstyle BCLN}}
 \def\arccosh{\,{\rm arccosh}\,}
 \def\arcsinh{\,{\rm arcsinh}\,}
\begin{document}
\title{Canonical and quantum FRW cosmological solutions in M-theory\thanks{This work is supported in part by funds provided by the U.S.\
Department of Energy (D.O.E.) under cooperative research agreement
DE-FC02-94ER40818.}}
\author{Marco Cavagli\`a~$^{(a,b,c)}~$\footnote{Post office address:
Massachusetts Institute of Technology,
Marco Cavagli\`a 6-408, 77 Massachusetts Avenue,
Cambridge MA 02139-4307, USA. E-mail: cavaglia@mitlns.mit.edu} , 
Paulo Vargas Moniz~$^{(b,d)}~$\footnote{E-mail: pmoniz@mercury.ubi.pt, 
web page: http://www.dfis.ubi.pt/\~{\null}pmoniz}\\
\small $^{(a)}$ Center for Theoretical Physics,
Laboratory for Nuclear Science and Department of Physics,\\
\small Massachusetts Institute of Technology,
77 Massachusetts Avenue, Cambridge MA 02139, USA\\
\small $^{(b)}$ Departamento de F{\'\i}sica,
Universidade da Beira Interior,\\
\small Rua Marqu{\^e}s d'{\'A}vila e Bolama,
6200 Covilh{\~a}, Portugal\\
\small $^{(c)}$ INFN, Sede di Presidenza, Roma, Italia\\
\small $^{(d)}$ CENTRA, IST, Av. Rovisco Pais, 1049 
Lisboa Codex, Portugal}
\date{(MIT-CTP-3037, hep-th/0010280. \today)}
\maketitle
\begin{abstract}
We present the canonical and quantum cosmological investigation of a spatially
flat,  four-di\-men\-sio\-nal Friedmann-Robertson-Walker (FRW) model that is
derived from the M-theory effective action obtained originally by Billyard,
Coley, Lidsey and Nilsson (BCLN). The analysis makes use of two sets of
canonical variables, the Shanmugadhasan gauge invariant canonical variables and
the ``hybrid'' variables which diagonalise the Hamiltonian. We find the
observables and discuss in detail the phase space of the classical theory. In
particular, a region of the phase space exists that describes a
four-dimensional FRW spacetime first contracting  from a strong coupling regime
and then expanding to  a weak coupling regime, while the internal space ever
contracts. We find the quantum solutions of the model and obtain the positive
norm Hilbert space of states. Finally, the correspondence between wave
functions and classical solutions is outlined.
\end{abstract}
\section{Introduction\label{intro}}
The search for a theory of quantum gravity constitutes one of the uppermost
challenges in theoretical high energy physics. The need for quantum gravity
finds its roots within Einstein general relativity. Powerful general theorems
\cite{EH} imply that our universe must have started from an initially singular
state with infinite curvature. In such circumstances, where the laws of
classical physics break down, it is unclear how any boundary conditions
necessary for a description of a dynamical system could have been imposed at
the initial singularity. Quantum corrections could induce a modification of
classical general relativity and strongly influence the evolution of the very
early universe \cite{QG}.

In the last two decades superstring theory \cite{strings} has emerged as a
successful candidate for the theory of quantum gravity. In cosmology, most of
the modifications to general relativity induced by superstring theory are
originated by the inclusion of the dilaton, axion and various moduli fields,
together with higher curvature terms that are present in the low-energy
effective actions. Each of these novel ingredients do indeed lead to new
cosmological solutions. A remarkable example is given by the so-called pre-big
bang cosmologies \cite{prebb} that are derived by the low-energy effective
string action. Different branches of the solution are related by time
reflection and internal transformations -- $O(d,d)$ and, in particular, scale
factor duality --  that follow from the $T$-duality property of the full
superstring theory \cite{duality}. According to the pre-big bang scenario the
universe evolves  from a weak coupled string vacuum state to a
radiation-dominated and then matter-dominated FRW geometry through a region of
strong coupling and large curvature. Although the pre-big bang model has not
yet proven able to solve all its pitfalls, such as the existence of a singular
boundary that separates the pre- and post-big bang branches \cite{exit}, it
provides a good starting point to investigate high-energy cosmology.

Recently, it has become evident that the five consistent, anomaly free,
perturbative formulations of ten-dimensional superstring theories are connected
by a web of duality transformations \cite{strings} and constitute special
points of a large, multi-dimensional moduli space of a more fundamental
(non-perturbative) theory, designated as M-theory \cite{M-th}. Quite
interestingly, it happens that another point of the M-theory moduli space
corresponds to eleven-dimensional supergravity, the low-energy limit of
M-theory \cite{M-th,Townsend}. Assuming that M-theory is the ultimate theory of
quantum gravity, it is natural to begin exploring its cosmological
implications. Although our understanding of M-theory is still incomplete,
there are hopes that some of the obstacles of dilaton driven inflation in
string theory could be overcome within the new theory. The underlying idea is
to investigate the dynamics at the extreme weak- and strong-coupling regimes of
superstring theory from a M-theory perspective, where the existence of eleven
dimensions seems mandatory. 

Several different approaches to M-theory cosmology have already been explored
in the literature \cite{LOW},\cite{RS},\cite{DH}-\cite{BCLN}. In the framework
of the Ho{\v r}ava-Witten model \cite{HW}, M-theory and cosmology have been
combined in the works of Lukas, Ovrut and Waldram \cite{LOW}. A somewhat
related line of research is the brane world by Randall and Sundrum \cite{RS},
where our four-dimensional universe emerges as the world volume of a
three-brane in a higher-dimensional spacetime. From a different point of view,
Damour and Henneaux have investigated chaotic models \cite{DH} and Lu,
Maharana, Mukherji and Pope have discussed classical and quantum M-theory
models with homogeneous graviton, dilaton and antisymmetric tensor field
strengths \cite{LMMP}. Different classes of cosmological solutions that reduce
to solutions of string dilaton gravity have been discussed in Ref.\
\cite{cosm}-\cite{KKO}. In particular, a global analysis of four-dimensional
cosmologies derived from M-theory and type $IIA$ superstring theory has been
presented by BCLN \cite{BCLN}. Using the theory of dynamical systems to
determine  the qualitative behaviour of the solutions, the authors find that
form-fields associated with the NS/NS and R/R sectors play a rather crucial
role in determining the dynamical behaviour of the solutions: the R/R fields
cause the universe to collapse but the NS/NS fields, such as the axion, have
opposite effect. Quite interestingly, for flat FRW models the boundary of the
classical physical phase space is a set of invariant submanifolds, where either
the axion field is trivial or the four-form field strength is dynamically
unimportant. This interplay leads to important consequences, as the orbits in
the phase space are dominated by the dynamics associated with one, or the
other, or both invariant submanifolds in sequence, shadowing trajectories in
the invariant submanifold \cite{BCLN}. 

In this paper we discuss the main scenario introduced by BCLN from a canonical
perspective. This approach allows us to analyse in depth the physical
properties of the classical solutions and to obtain a consistent quantum
description of the model. We consider the bosonic sector of eleven-dimensional
supergravity which consists of a graviton and an antisymmetric three-form
potential. The theory is compactified to four dimensions by assuming a geometry
of the form $M^4 \times T^6 \times S^1$, where $T^6$ is a six-dimensional torus
and $M^4$ corresponds to a spatially flat FRW spacetime. The effective theory
in four dimensions bears a dilaton $\phi$, a modulus field $\beta$ identifying
the internal space, a pseudo-scalar axion field $\sigma$ and a potential term
induced by a four-form field. [See Eq.\ (\ref{b5}) below.] A brief derivation
of the previous steps is presented in Section 2, where the two invariant
submanifolds are identified. In Section 3 we analyse the first invariant
submanifold, where the four-form field is negligible and the axion field
dominates. In particular, in Subsection 3.1 we discuss the parameter space of
the classical theory and in Subsection 3.2 we find the Hilbert space of states
of the quantum theory. This programme is performed by using two sets of
canonical variables, the so-called Shanmugadhasan variables \cite{Shan} 
(forming a maximal set of gauge invariant variables)  and the ``hybrid"
variables that diagonalise the Hamiltonian. In Section 4, we repeat the
analysis of Section 3 for the invariant manifold with negligible axion field
and dominant four-form field. Finally, our conclusions are presented in Section
5. Two appendixes contain technical details and complete the paper.
\section{M-theory cosmology\label{cosmology}}
In this section we closely follow Ref.\ \cite{BCLN} and derive the
four-dimensional minisuperspace effective actions that will be used to discuss
the dynamics of M-theory cosmology. 

The bosonic sector of eleven-dimensional supergravity action $S^{(11)}$ is
\ba
&&S^{(11)}=\int d^{11}X \sqrt{-g^{(11)}}\left[R^{(11)}(g^{(11)}_{ab})
-{1\over 48}F_{a_1\dots a_4}F^{a_1\dots a_4}\right.\nonumber\\
&&\qquad\qquad\left. -{1\over 12^4\sqrt{-g^{(11)}}}\epsilon^{a_1 \dots a_3 
b_1 \dots b_4 c_1 \dots c_4}
A_{a_1 \dots a_3}F_{b_1 \dots b_4}F_{c_1 \dots c_4} \right]\,,
\label{b1}\ea
where $a_i,b_i,c_i=0\dots 10$, $F_{a_1 \dots a_4}=4\partial_{[a_1}A_{a_2 \dots
a_4]}$ is the four-form field strength of the antisymmetric three-form
potential $A_{a_1 \dots a_3}$, and $g^{(11)}$ denotes the determinant of the
eleven-dimensional metric $g^{(11)}_{ab}$. Equation (\ref{b1}) describes the
low-energy limit of M-theory \cite{M-th,Townsend}.

The four-dimensional effective action is derived from Eq.\ (\ref{b1}) by a
sequence of two Kaluza-Klein compactifications, on a circle $S^1$ with radius
$R_{S^1}=e^{\Phi_{10}/3}$ and on an isotropic six-torus with volume
$V_{T^6}=e^{6\beta}$, respectively. The eleven-dimensional metric is
\be
ds^2_{(11)}\equiv g^{(11)}_{ab}dX^adX^b=e^{-\Phi_{10}/3}
[g_{\mu\nu} dx^{\mu} dx^{\nu}+e^{2\beta} 
\delta_{ij} dx^idx^j]+e^{2\Phi_{10}/3}dX_{10}^2\,,
\label{b2}
\ee
where $\mu,\nu=0\dots 3$ and $i,j=4\dots 9$. The first compactification gives
the ten-dimensional effective action for the massless type IIA superstring
\cite{M-th,compactifI}
\ba
&&S^{(10)}=\int d^{10}x \sqrt{-g^{(s)}} \left[e^{-\Phi_{10}}
\left(R^{(10)}(g^{(s)}_{mn})
+\left(\nabla \Phi_{10} \right)^2-{1\over 12}H_{mnp}H^{mnp}
\right)\right.\nonumber\\
&&\qquad\qquad\left. -\frac{1}{48}F_{mnpq}F^{mnpq}-{1\over 384\sqrt{-g^{(11)}}}
\epsilon^{m_1 m_2 n_1 \dots n_4 p_1 \dots p_4}
B_{m_1 m_2}F_{n_1 \dots n_4}F_{p_1 \dots p_4}\right]\,,
\label{b3}
\ea
where $g^{(s)}_{mn}=(g_{\mu\nu},e^{2\beta}\delta_{ij})$, and $H_{mnp}$ and
$F_{mnpq}$ are the field strengths of the potentials $B_{np}$ and $A_{npq}$,
respectively. Assuming that the only non-trivial components of the form fields
are those associated with $M^4$ the second compactification gives the effective
four-dimensional action \cite{BCLN}
\be 
S=\int d^4x\sqrt{-g}\left[e^{-\Phi_4}\left(R^{(4)}(g_{\mu\nu})
+\left(\nabla\Phi_4\right)^2-6\left(\nabla\beta\right)^2-{1\over 12} 
H_{\mu\nu\lambda}H^{\mu\nu\lambda}\right)-{1\over 48}e^{6\beta} 
F_{\mu\nu\lambda\kappa}F^{\mu\nu\lambda\kappa}\right]\,,
\label{b4}
\ee
where the four-dimensional dilaton field is $\Phi_{4}=\Phi_{10}-6\beta$.
Finally, solving the field equation for the four-form $F^{\mu\nu\lambda\kappa}$
and dualizing the three-form $H^{\mu\nu\lambda}$, Eq.\ (\ref{b4}) can be cast
in the form
\be
S=\int d^4x\sqrt{-g}\left[e^{-\Phi_4}\left(R^{(4)}(g_{\mu\nu})+\left( 
\nabla\Phi_4\right)^2-6\left(\nabla\beta\right)^2-{1\over 2} 
e^{2\Phi_4}\left(\nabla\sigma \right)^2 \right) 
-{1\over 2}Q^2 e^{-6\beta}\right]\,,
\label{b5}
\ee
where $\sigma$ is the pseudo-scalar axion field dual to the three-form,
$H^{\mu\nu\lambda}=e^{\Phi_4}\epsilon^{\mu\nu\lambda\kappa}
\nabla_{\kappa}\sigma$, and $F^{\mu\nu\lambda\kappa}=Qe^{-6\beta}
\epsilon^{\mu\nu\lambda\kappa}$. Equation (\ref{b5}) is our starting
point to investigate classical and quantum four-dimensional M-theory
cosmology. 

Since we are interested in homogeneous and isotropic four-dimensional
cosmologies the ansatz for the four-dimensional section of the
(string frame) metric (\ref{b2}) is
\be
ds^2_{(4)}\equiv
g_{\mu\nu}dx^\mu dx^\nu=-N^2(t)dt^2+e^{2\alpha(t)}d\Omega_{3k}\,,\qquad
N(t)>0
\label{c1}
\ee
where $d\Omega_{3k}$ is the maximally symmetric three-dimensional unit metric
with curvature $k=0,\pm 1$, respectively. By substituting Eq.\ (\ref{c1}) in
Eq.\ (\ref{b5}) and requiring that the modulus field $\beta$, the dilaton
$\Phi_4$, and the axion $\sigma$ depend only on $t$, the action (\ref{b5})
becomes
\be
S=\int
dt\,\left[{1\over\mu}\left(3\dot\alpha^2-\dot\phi^2
+6\dot\beta^2+{1\over 2}\dot\sigma^2e^{2(3\alpha+\phi)}\right)+
\mu\left(6k e^{-2(\alpha+\phi)}-{1\over
2}Q^2e^{3\alpha-\phi-6\beta}\right)\right]\,,
\label{c2}
\ee
where we have defined the ``shifted dilaton'' field
\be
\phi=\Phi_4-3\alpha\,,
\label{c3}
\ee
and the Lagrange multiplier
\be
\mu(t)=Ne^{\phi}>0\,.\label{mu}
\label{c4}
\ee
The dynamics of the action (\ref{c2}) has been discussed qualitatively in Ref.\
\cite{BCLN}. Here we restrict attention to spatially flat spacetimes ($k=0$)
and discuss in detail two subcases of Eq.\ (\ref{c2}) that turn out to be
completely integrable:
\begin{itemize}
\item[\bf I] The invariant submanifold
\be
S_I=\int
dt\,\left[{1\over\mu}\left(3\dot\alpha^2-\dot\phi^2
+6\dot\beta^2+{1\over 2}\dot\sigma^2e^{2(3\alpha+\phi)}\right)\right]\,,
\label{c5}\ee
which is obtained from Eq.\ (\ref{c2}) by setting $Q=0$ and $k=0$. This case
describes spatially flat low-energy M-theory cosmology with negligible
four-form $F^{\mu\nu\lambda\kappa}$. The general solution for this model
(including spatially curved models which are not discussed here) was first
discussed by Copeland, Lahiri and Wands in Ref.\ \cite{CLW}.
\item[\bf II] The invariant submanifold
\be
S_{II}=\int
dt\,\left[{1\over\mu}\left(3\dot\alpha^2-\dot\phi^2
+6\dot\beta^2\right)-
{1\over 2}\mu\,Q^2e^{3\alpha-\phi-6\beta}\right]\,.
\label{c6}\ee
which is obtained from Eq.\ (\ref{c2}) by setting $\sigma=constant$ and $k=0$.
This case describes spatially flat low-energy M-theory cosmology with trivial
axion field.
\end{itemize}
The intersection of the two invariant submanifolds I and II further identifies
the invariant submanifold
\be
S_{I\cap II}=\int
dt\,\left[{1\over\mu}\left(3\dot\alpha^2-\dot\phi^2
+6\dot\beta^2\right)\right]\,.
\label{c7}\ee
In the following two sections we will discuss the classical and quantum
dynamics of the invariant submanifolds I and II, respectively. The degenerate
case $S_{I\cap II}$ will be outlined at the end of Section \ref{caseII}.
\section{Invariant submanifold I\label{caseI}}
Equation (\ref{c5}) can be cast in the canonical form
\be
S_I=\int dt \left[\dot\alpha p_\alpha+\dot\phi p_\phi+\dot\beta
p_\beta+\dot\sigma p_\sigma-{\cal H}\right]\,,
\label{d1}
\ee
where the Hamiltonian is
\be
{\cal H}=\mu H\,,\qquad
H={1\over 24}\left(2p^2_\alpha-6p^2_\phi+p^2_\beta
+12p^2_\sigma e^{-2(3\alpha+\phi)}\right)\,.\label{HII}
\label{d2}
\ee
As is expected for a time-reparametrization invariant system, the total
Hamiltonian ${\cal H}$ is proportional to the non-dynamical variable $\mu$
\cite{HRT,HT}. The latter enforces the constraint 
\be
H=0\qquad\to\qquad 
2p_\alpha^2-6p^2_\phi+p^2_\beta+12p^2_\sigma e^{-2(3\alpha+\phi)}=0\,.
\label{constraintII}
\ee
The canonical equations of motion are
\be
\begin{array}{llll}
\displaystyle
\dot\alpha={p_\alpha\over 6}\,,\quad
&\displaystyle
\dot\phi=-{p_\phi\over 2}\,,\quad
&\displaystyle
\dot\beta={p_\beta\over 12}\,,\quad
&\displaystyle
\dot\sigma=p_\sigma e^{-2(3\alpha+\phi)}\,,\\\\
\displaystyle
\dot p_\alpha=3p_\sigma^2e^{-2(3\alpha+\phi)}\,,\quad
&\displaystyle
\dot p_\phi=p_\sigma^2e^{-2(3\alpha+\phi)}\,,\quad
&\displaystyle
\dot p_\beta=0\,,\quad
&\displaystyle
\dot p_\sigma=0\,,
\end{array}\label{eqsII}
\ee
where the dots represent differentiation w.r.t.\ gauge parameter
\be
\tau(t)=\int_{t_0}^t \mu(t') dt'\,,\qquad t>t_0\,,
\label{d5}
\ee
where $t_0$ is an arbitrary constant. Note that since $\mu$ is positive defined
$\tau(t)$ is a monotone increasing function.

Different choices of $\mu$ correspond to different choices of time. We have the
following interesting time parameters:
\begin{itemize}
\item[{\it i)}] Cosmic proper time $t_c$. The cosmic proper time is defined by
the condition $N=1$. It is obtained by choosing
\be
\mu(\tau)\equiv\left[{dt_c\over d\tau}\right]^{-1}=e^\phi\,,
\label{d6}
\ee
which leads to
\be
t_c(\tau)=\int_{-\infty}^\tau d\tau\, e^{-\phi(\tau')}\,,
\label{d7}
\ee
where we have chosen the arbitrary constant so that $t(-\infty)=0$.
\item[{\it ii)}] Gauge proper time $t_g$. The gauge proper time is defined by
the condition $\mu=1$. This leads to
\be
t_g(\tau)=\tau-\tau_0\,.
\label{d8}
\ee
This is the natural time parameter to discuss the classical dynamics and the
quantization of the model.
\item[{\it iii)}] The BCLN time variable $\eta$. This is defined by the condition
\be
{d\eta\over dt_c}\equiv e^{(3\alpha+\phi-6\beta)/2}\,,
\label{eb}
\ee
or, using Eq.\ (\ref{d7}),
\be
{d\eta\over d\tau}=e^{(3\alpha-\phi-6\beta)/2}\,,
\label{etab}
\ee
\item[{\it iv)}] The BCLN dimensionless time variable $\Tb$
\be
{d\Tb\over d\eta}={d\phi\over d\eta}\,,\quad\to\quad \Tb-\Tb{}_0=\phi\,.
\label{taub}
\ee
\end{itemize}
The choices {\it iii)} and {\it iv)} are useful to compare the notations used in
this paper to those of Ref.\ \cite{BCLN}. This is done in Appendix A.
\subsection{Classical solutions}
The off-shell solution of the canonical equations is\footnote{Here and throughout
the section we assume $\kappa\not=0$ ($p_\sigma\not=0$). The case $\kappa=0$
corresponds to the degenerate case $S_{I\cap II}$ and will be discussed at the
end of Section \ref{caseII}.}
\be
\begin{array}{lll}
\alpha&=&\displaystyle\alpha_0+{1\over 2}\ln
\left[\cosh\left(\kappa(\tau-\tau_0)\right)\right]
-\xi(\tau-\tau_0)\,,\\\\
p_\alpha&=&\displaystyle 3\kappa\tanh\left[\kappa(\tau-\tau_0)\right]-6\xi\,,\\\\
\phi&=&\displaystyle\phi_0-{1\over 2}\ln
\left[\cosh\left(\kappa(\tau-\tau_0)\right)\right]+
3\xi(\tau-\tau_0)\,,\label{phi}\\\nonumber\\
p_\phi&=&\displaystyle\kappa\tanh\left[\kappa(\tau-\tau_0)\right]-6\xi\,,\\\\
\beta&=&\displaystyle\beta_0+{p_\beta\over 12}(\tau-\tau_0)\,,\\\\
p_\beta&=&{\rm constant}\,,\\\\
\sigma&=&\displaystyle\sigma_0+{\kappa\over
p_\sigma}\tanh\left[\kappa(\tau-\tau_0)\right]\\\\
p_\sigma&=&{\rm constant}\,,
\end{array}
\label{solII}
\ee
where $\alpha_0$, $\phi_0$, $\beta_0$, $\sigma_0$ and $\tau_0$ are constants of
integration, 
\be
\kappa^2-12\xi^2+{p_\beta^2\over 12}=2H\,,\qquad\kappa\not=0\,,
\label{kappaII}
\ee
and (we choose $\kappa>0$ for simplicity)
\be
3\alpha_0+\phi_0=\ln\left({|p_\sigma|\over\kappa}\right)\,.
\ee
On the (physical) shell $\xi=0$ is a degenerate trivial configuration of the
system because it implies $\kappa=0$ and $p_\beta=0$. 

Since the model is integrable we can find a maximal set of gauge invariant
observables. The system is invariant under reparametrizations of time so we
expect six physical gauge invariant observables. (Indeed, the system possesses
four canonical degrees of freedom and thus the general solution is determined
by eight constant of motion: the constraint $H$, the initial value of the gauge
parameter $\tau$, and six other constant quantities which are the observables
of the system.) Considering the off-shell solution a possible choice is
\be
\displaystyle
\begin{array}{lll}
\displaystyle
U=\alpha+\phi+{3 p_\phi-p_\alpha\over 6\kappa}
\arccosh\left({\kappa\over |p_\sigma|}e^{\phi+3\alpha}\right)\,,&&
\displaystyle
V={1\over 2}\left(3p_\phi-p_\alpha\right)\,,\\\\
\displaystyle
X=\beta-{p_\beta\over 12\kappa}
\arccosh\left({\kappa\over |p_\sigma|} e^{\phi+3\alpha}\right)\,,&&
\displaystyle
W=p_\beta\,,\\\\
\displaystyle Y=\sigma-{p_\alpha-p_\phi\over 2p_\sigma}\,,&&
\displaystyle Z=p_\sigma\,,
\end{array}\label{obsII}
\ee
where
\be
\kappa=\left[{1\over 4}\left(p_\alpha-p_\phi\right)^2+p_\sigma^2
e^{-2(3\alpha+\phi)}\right]^{1/2}\,.
\ee
The quantities (\ref{obsII}), being gauge invariant, have vanishing Poisson
brackets with the Hamiltonian (\ref{HII}) and have been chosen such that the
only nonvanishing Poisson brackets are
\be
\begin{array}{lllll}
\left[U,V\right]_P=1\,,&&\left[X,W\right]_P=1\,,
&&\left[Y,Z\right]_P=1\,.
\end{array}\label{canonicalPBII}
\ee
The set of gauge invariant quantities (\ref{obsII}) can be completed by the
Hamiltonian $H$ and by its canonically conjugate ($[T,H]_P=1$)
\be
T={1\over\kappa}
\arccosh\left({\kappa\over |p_\sigma|} e^{\phi+3\alpha}\right)\label{TII}
\ee
to obtain a maximal set of canonical variables (also called Shanmugadhasan
variables \cite{Shan}). Note that the quantity $T$ has been chosen to have
vanishing Poisson brackets with the observables (\ref{obsII}). Since
$[T,H]_P=1$, $T$ transforms linearly under gauge transformations generated by
the constraint. This will be useful in the following.

For sake of completeness, let us write the relation between the gauge invariant
observables and the constants of motion in Eqs.\ (\ref{solII}). Using Eqs.\
(\ref{HII}), (\ref{obsII}) and (\ref{TII}), after a bit of algebra we find
\be
\begin{array}{c}
\displaystyle \alpha_0={1\over 2}\ln{|p_\sigma|\over\kappa}-
{1\over 2}U\,,\quad
\phi_0=-{1\over 2}\ln{|p_\sigma|\over\kappa}+
{3\over 2}U\,,\quad
\beta_0=X\,,\quad
\sigma_0=Y\,,\\\\
\displaystyle \xi=-{1\over 6}V\,,\quad
p_\beta=W\,,\quad
p_\sigma=Z\,,\quad
\kappa=\left[{1\over 12}(4V^2-W^2)+2H\right]^{1/2}\,.
\end{array}\label{fieldsII}
\ee

Another useful canonical chart is formed by the hybrid variables that
diagonalise the constraint (\ref{constraintII}). Although the hybrid variables
are not (all) gauge invariant they allow to fix a global gauge and quantize
exactly the system. The hybrid variables $(a,b,c,\sigma)$ are defined by the
canonical transformation
\be
\displaystyle
\begin{array}{lll}
\displaystyle
a=\phi+3\alpha\,,\quad
&\displaystyle
b=\sqrt{3}(\phi+\alpha)\,,\quad
&\displaystyle
c=2\sqrt{3}\beta\,,\\\\
\displaystyle
p_a={1\over 2}\left(p_\alpha-p_\phi\right)\,,\quad
&\displaystyle
p_b={1\over 2\sqrt{3}}\left(3p_\phi-p_\alpha\right)\,,\quad
&\displaystyle
p_c={1\over 2\sqrt{3}}p_\beta\,.
\end{array}\label{hybridII}
\ee
Note that $a$ coincides with the four-dimensional dilaton field $\Phi_4$. Using
the hybrid variables the constraint (\ref{constraintII}) reads (we have divided
by a factor $12$)
\be
p_a^2-p^2_b+p^2_c+p^2_\sigma e^{-2a}=0\,.
\label{constraint-hybII}
\ee
The gauge invariant canonical variables are related to the hybrid
variables by the canonical transformation 
\be
\displaystyle
\begin{array}{ll}
\displaystyle
U={1\over\sqrt{3}}\left[b+{p_b\over\kappa}
\arccosh\left({\kappa\over |p_\sigma|}e^{a}\right)\right]\,,\quad
&\displaystyle
V=\sqrt{3}p_b\,,\\\\
\displaystyle
X={1\over 2\sqrt{3}}\left[c-{p_c\over\kappa}
\arccosh\left({\kappa\over |p_\sigma|} e^{a}\right)\right]\,,\quad
&\displaystyle
W=2\sqrt{3}p_c\,,\\\\
\displaystyle
Y=\sigma-{p_a\over p_\sigma}\,,\quad
&\displaystyle
Z=p_\sigma\,,\\\\
\displaystyle
T={1\over\kappa}
\arccosh\left({\kappa\over |p_\sigma|} e^{a}\right)\,,\quad
&\displaystyle
H={1\over 2}\left(p_a^2-p^2_b+p^2_c+p^2_\sigma e^{-2a}\right)\,,
\end{array}\label{gauge-to-hybII}
\ee
where $\kappa=\sqrt{p_a^2+p_\sigma^2e^{-2a}}$. The canonical transformation
above is generated by the generating function of the first kind \cite{Goldstein}
\be
F_1(a,b,c,\sigma;U,X,Y,T)={1\over
2T}\left[(\sqrt{3}U-b)^2-(2\sqrt{3}X-c)^2-
\arccosh^2\sqrt{1+e^{2a}(Y-\sigma)^2}\right]\,.
\label{gfunctionII}
\ee

Let us now discuss the behaviour of the classical solution. The on-shell
classical solution is determined by six physical parameters. From the equations
of motion we see that five of them ($\alpha_0$, $\phi_0$, $\beta_0$,
$\sigma_0$, and $\tau_0$) give initial conditions for the canonical variables
and do not influence the qualitative behaviour of the solution. Therefore, the
qualitative dynamics of the model is determined by a two-dimensional parameter
space described by two coordinates, for instance $\kappa$ and $\xi$. Using
$\kappa$ and $\xi$ as free parameters, from Eq.\ (\ref{kappaII}) it follows
that $p_\beta$ is (on-shell)
\be
p_\beta=\pm{2\sqrt{3}}\sqrt{12\xi^2-\kappa^2}\,.
\label{hypI}
\ee
The sign of $p_\beta$ determines the dynamical behaviour of the internal
six-torus space. In fact, from the solution of the equations of motion one
obtains the scale factor of the internal space (we set $\tau_0=0$ for
simplicity) 
\be
R_{T^6}\equiv e^\beta=e^{\beta_0}\cdot e^{p_\beta\tau/12}\,.
\ee
Physically, we want compactification at late (cosmic) times.\footnote{A
successful physical model ultimately requires that the moduli fields are
stabilized as well. Stabilization of the internal space does not occur in the
models under consideration, where only a fraction of all the degrees of freedom
present in Eq.\ (\ref{b1}) are considered, with exception of the (fine-tuned)
case $p_\beta=0$ (see below). Hopefully, the inclusion of more degrees of
freedom will provide a mechanism for stabilization of extra-dimensions at late
times.} Compactification of the six-torus space is achieved for $p_\beta\le
0$. Indeed, for negative values of $p_\beta$ $\lim_{\tau\to\infty} R_{T^6}=0$,
i.e., the internal space shrinks to zero for large values of the gauge proper
time, while decompatifying for $\tau\to-\infty$ when the strong coupling region
of the theory is approached. Since the relation between the cosmic proper time
and the gauge proper time is monotone [see Eq.\ (\ref{d7})] the dynamics in
$\tau$ is identical to the dynamics in $t_c$, and the internal space shrinks to
zero for large values of the cosmic proper time as well. Note that the gauge
proper time is defined in the interval $\tau\in]-\infty,\infty[$ whereas $t_c$
is defined only on the half line. The limiting value $p_\beta=0$ corresponds to
a constant (stable) internal space with radius $R_{T^6}=e^{\beta_0}$ and is
physically acceptable for sufficiently large negative values of $\beta_0$. In
the following we will restrict attention to nonpositive values of $p_\beta$,
the extension to $p_\beta>0$ being straightforward (see Fig.\ \ref{figI}).

At fixed $\kappa$ we distinguish three different dynamical behaviours of the
four-dimensional external space according to the value of $\xi$:
\begin{figure}
\centerline{\epsfxsize=4.0in \epsffile{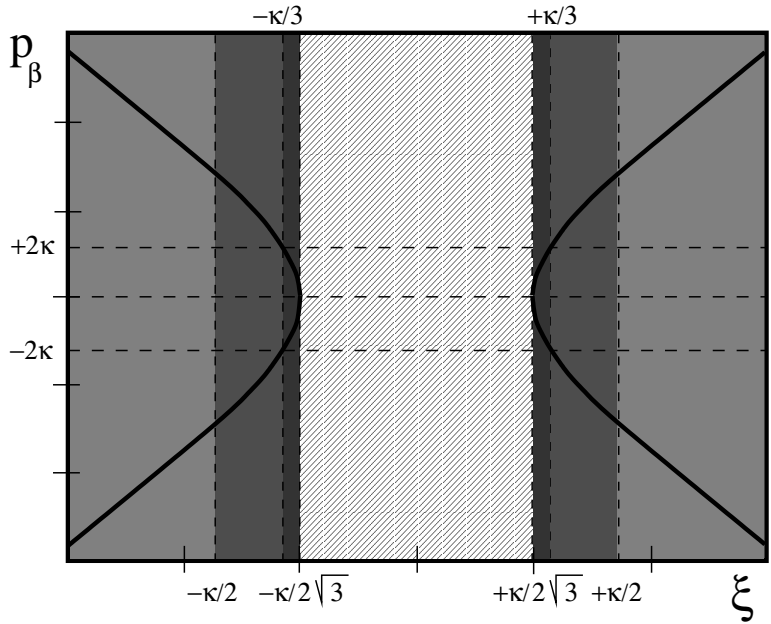}}
\caption{\small Parameter space for invariant submanifold I. The physical
points are represented by the two branches of the hyperbola (\ref{hypI}). The
different colored regions determine on the two branches of the hyperbola the
different physical cases described in the text.}
\label{figI}
\end{figure}
\begin{itemize} 
\item[{\it i)}] $\xi\le-\kappa/2$ ($p_\beta<-2\kappa$). In this case the
external scale factor ever expands while the internal scale factor shrinks from
infinity to zero. The Hubble parameter is always positive and vanishes
asymptotically at large times. In particular, for $\xi=-\kappa/2$ the external
space starts at $\tau=-\infty$ ($t_c=0$) with a finite nonzero scale factor and
vanishing Hubble parameter. For $\xi<-\kappa/2$ the external space starts with
a vanishing scale factor and infinite Hubble parameter, which is always
decreasing. $\tau=-\infty$ ($t_c=0$) is the strong coupling region where the
coupling constants of the theory, $g=\exp(\phi)$ and $g_{10}=\exp(\Phi_{10})$,
become infinite. Conversely, $\tau=\infty$ ($t_c=\infty$) is the weak region
coupling where both $g$ and $g_{10}$ vanish. $g$ and $g_{10}$ are always
decreasing. (See Fig.\ \ref{figIab}.) 
\begin{figure}
\centerline{\epsfysize=150pt \epsffile{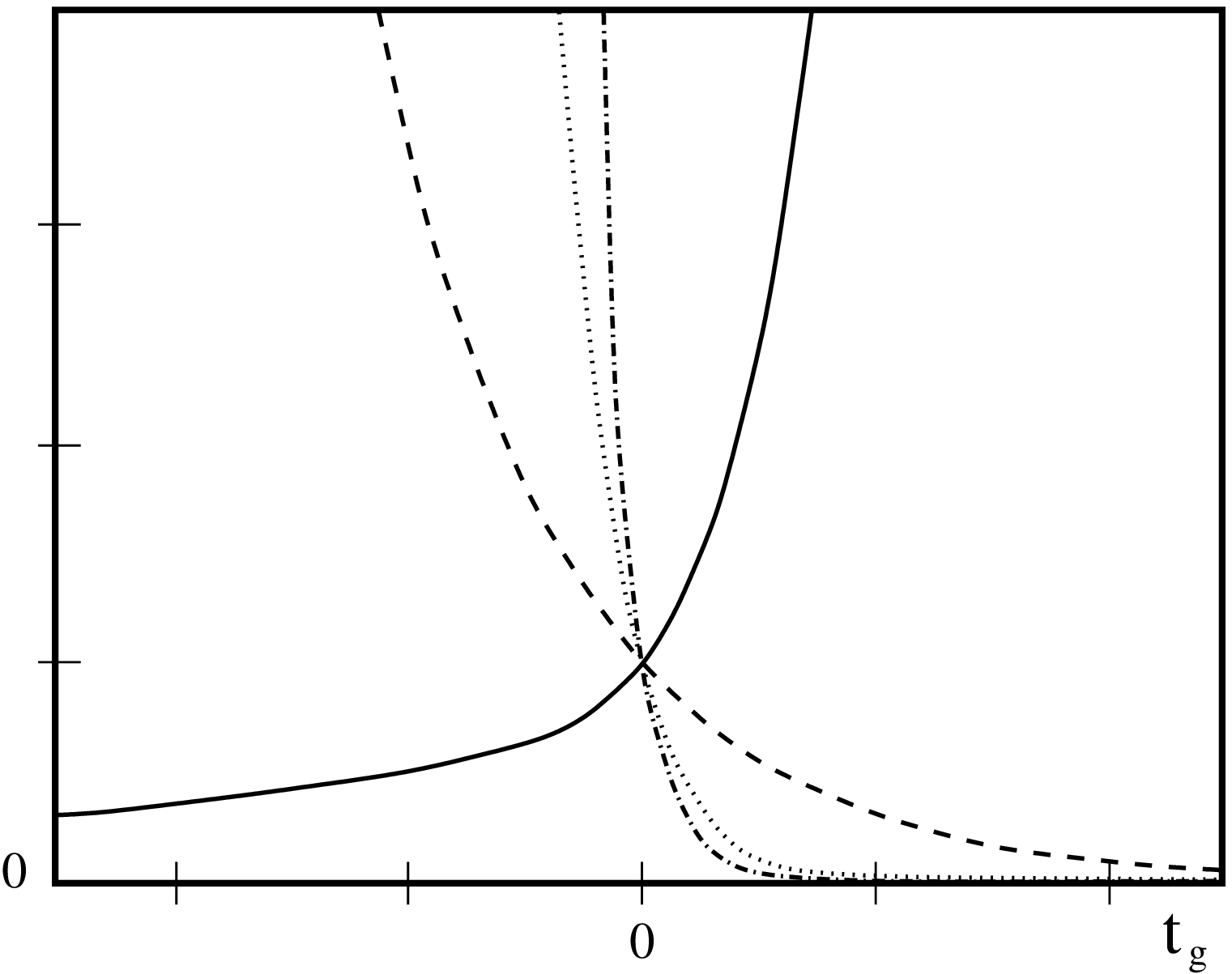}\qquad
\epsfysize=150pt \epsffile{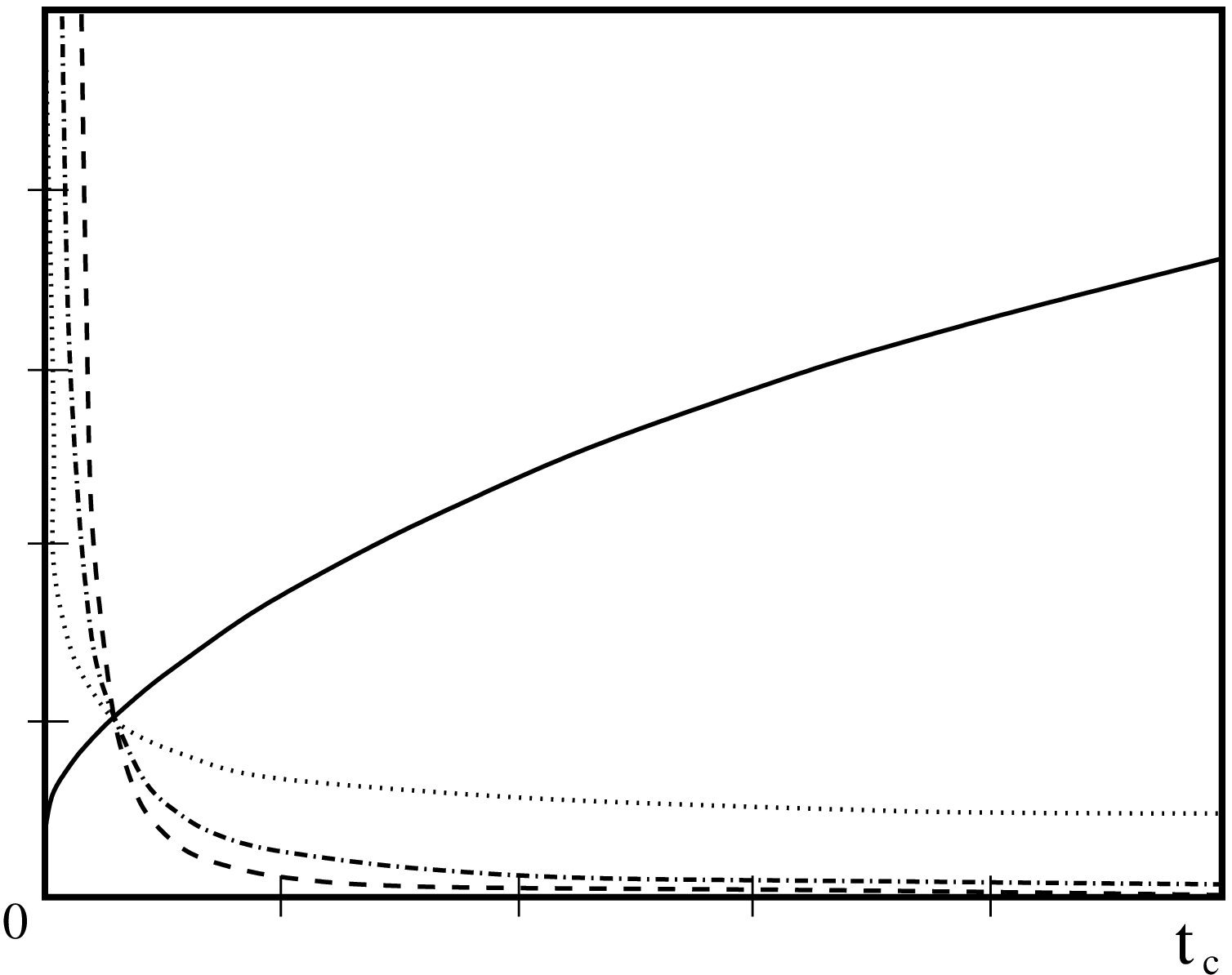}}
\caption{\small Graphical representation of the dynamics of the invariant
submanifold I for $\xi<-\kappa/2$ as function of the gauge proper time $t_g$
and of the cosmic proper time $t_c$. [Here and in the following figures we set
$\alpha_0=\phi_0=\beta_0=0$ for simplicity.] The solid and dashed lines
describe the evolution of the radius of the external spacetime and of the
internal space, respectively. The dotted and dash-dotted lines represent
the evolution of the coupling constants $g$ and $g_{10}$, respectively.}
\label{figIab}
\end{figure}
\item[{\it ii)}] $\xi\ge\kappa/2$ ($p_\beta<-2\kappa$). In this case both the
external scale factor and the internal scale factor ever contract. The Hubble
parameter is always negative and asymptotically vanishing at small times
($\tau=-\infty$). In particular, for $\xi=\kappa/2$ the external space ends at
$\tau=\infty$ ($t_c=0$) with a finite nonzero scale factor and vanishing Hubble
parameter. For $\xi>\kappa/2$ the external space ends with a vanishing scale
factor and infinite Hubble parameter.  $g$ increases from zero at
$\tau=-\infty$ ($t_c=-\infty$) to infinity at $\tau=\infty$ ($t_c=0$).
Conversely, $g_{10}$ decreases from infinity to zero. 
\item[{\it iii)}] $-\kappa/2 < \xi < -\kappa/2\sqrt{3}$ and $\kappa/2\sqrt{3}
<\xi<\kappa/2$. In this case the external scale factor first contracts then
expands, bouncing from infinity to infinity. In particular, for 
\begin{itemize}
\item[a)] $-\kappa/2<\xi\le-\kappa/3$ ($p_\beta<-2\kappa$) the internal scale
factor shrinks from infinity to zero. The Hubble parameter starts with infinite
negative value, becomes positive and then decreases to zero after having
reached a positive maximum. $g$ and $g_{10}$ decrease from infinity to zero
[$g_{10}$ to a finite nonzero positive value for the limiting value
$\xi=-\kappa/3$] (see Fig.\ \ref{figIcd});
\begin{figure}
\centerline{\epsfysize=150pt \epsffile{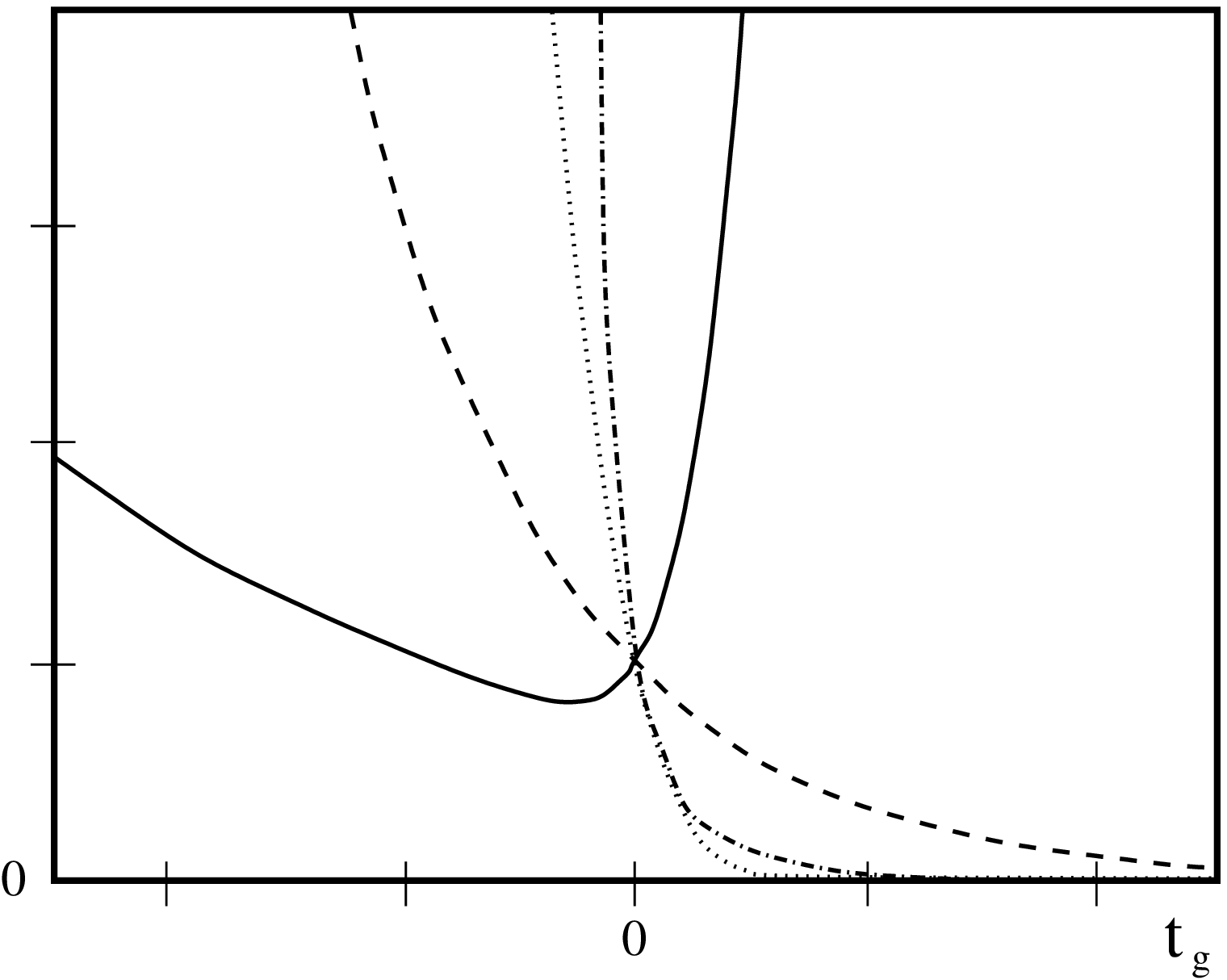}\qquad
\epsfysize=150pt \epsffile{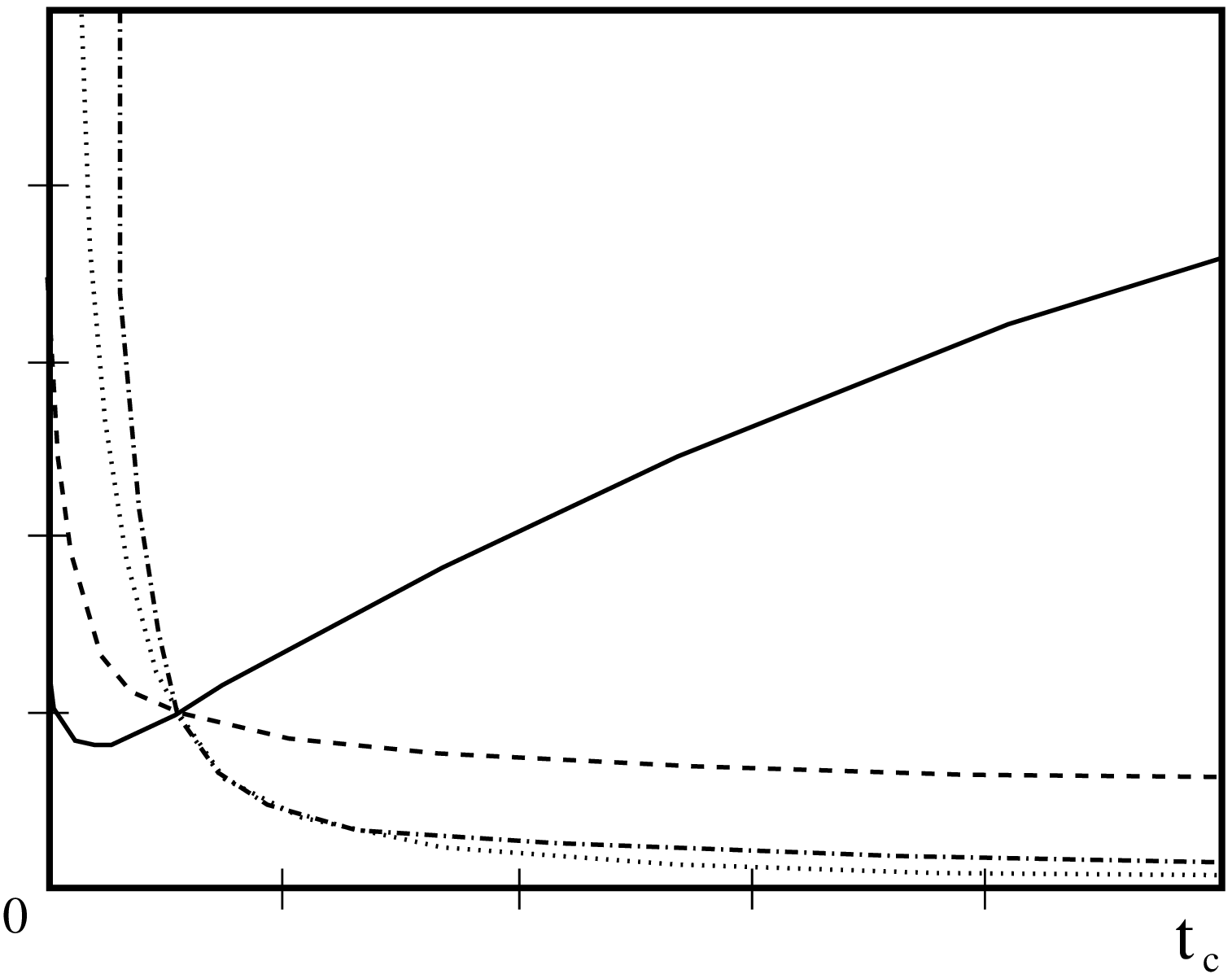}}
\caption{\small Graphical representation of the dynamics of the invariant
submanifold I for $-\kappa/2<\xi<-\kappa/3$ as function of the gauge proper 
time $t_g$ and of the cosmic proper time $t_c$. The internal scale factor and
the coupling constant $g$ and $g_{10}$ are ever decreasing and vanish for
large times. The external scale factor is large both at small and large times.
(See Fig.\ \ref{figIab} for the key to the plotted lines.)}
\label{figIcd}
\end{figure}
\item[b)] $-\kappa/3<\xi\le-\kappa/2\sqrt{3}$ the internal scale factor shrinks
from infinity to zero [$-\kappa/3<\xi<-\kappa/2\sqrt{3}$
($-2\kappa<p_\beta<0$)] or is constant [$\xi=-\kappa/2\sqrt{3}$ 
($p_\beta=0$)].The Hubble parameter starts with infinite negative value,
becomes positive and then decreases to zero after having reached a positive
maximum. $g$ decreases from infinity to zero. $g_{10}$ bounces from infinity
to infinity via a positive minimum;
\item[c)] $\kappa/2\sqrt{3}\le\xi<\kappa/3$ ($-2\kappa<p_\beta<0$). The
internal scale factor shrinks from infinity to zero
[$\kappa/2\sqrt{3}<\xi<\kappa/3$ ($-2\kappa<p_\beta<0$)] or is constant [$\xi
=\kappa/2\sqrt{3}$ ($p_\beta=0$)]. The Hubble parameter is first negative and
small, decreases to a negative minimum and then increases to infinity.  $g$
increases from zero to infinity.  $g_{10}$ bounces from infinity to infinity
via a positive minimum.
\item[d)] $\kappa/3\le\xi<\kappa/2$ ($p_\beta<-2\kappa$). The internal scale
factor shrinks from infinity to zero. The Hubble parameter is first negative
and small, decreases to a negative minimum and then increases to infinity. $g$
increases from zero to infinity. $g_{10}$ decreases from infinity to zero for
$\kappa/3<\xi<\kappa/2$ or to a finite nonzero positive value for the limiting
value $\xi=\kappa/3$.
\end{itemize}
\end{itemize}

Several comments are in order about the dynamics of the dilaton-axion-moduli
solution {\it i)-iii)}. First of all, we do not find a two-branch solution as
happens in the pre-big bang scenario \cite{prebb}. Indeed, the action
(\ref{c5}) (with $\dot\sigma\not=0$) is not invariant under scale factor
duality and moreover the presence of the axion leads to an effective potential 
[see Eq.\ (\ref{constraint-hybII})] with opposite sign to that required for the
pre-/post-big bang branches to exist. (See e.g.\ Gasperini, Maharana and
Veneziano in Ref.\ \cite{exit}.) Scenarios {\it i)} and {\it iii)} are most
attractive as far as a physical description of our universe is concerned.
According to {\it i)} the universe begins in a strong coupling region with
large internal space. Actually, both the coupling constants, $g$ and $g_{10}$,
and the radius of the internal six-torus are infinite at $t_c=0$. However, this
is not disturbing because for $t_c\to 0$ the spacetime curvature blows up. So
we are in the strong coupling regime of the theory and the low-energy
description of M-theory given by eleven-dimensional supergravity breaks down.
Hopefully, non-perturbative effects would cure the initial singularity. At
small times ($t_c\approx 0$) the expansion of the external spacetime is
characterized $H>>1$. Then the expansion slows down and at large times the
Hubble parameter vanishes, approaching a standard FRW behaviour. At small
cosmic times the scale factor of the external spacetime expands as
\be
a(t_c)\approx t_c^{(2\xi/\kappa+1)/(6\xi/\kappa+1)}\,.
\label{aext}
\ee
This behaviour coincides with that found by Copeland, Lahiri and Wands in Ref.\
\cite{CLW}. (See Eq.\ (4.43) and (4.46) with $\xi/\kappa=\pm
n\xi_\alpha/\Delta$.) For $\xi<-\kappa/2$ the exponent in Eq.\ (\ref{aext})
takes values in the interval $]0,1/3[$. This implies that we have no
inflationary expansion in the model. (Here and throughout the paper we follow
Ref.\ \cite{CLW} and define inflation as accelerated growth of the external
scale factor, namely $d^2a/dt^2_c>0$.) Inflation happens instead in case {\it
iii)}, where the external spacetime is first contracting and eventually
expanding. For this region of the parameter space the internal space is ever
decreasing. At small times both the internal space and the ten-dimensional
coupling constant are decompactified. At large times the eleventh dimension may
be compactified [cases a)-d)] or decompactified [cases b)-c)]. Among the former
cases, a) (Fig.\ \ref{figIcd}) is the most relevant from the physical point of
view since the Hubble parameter vanish asymptotically at large times.

In the next subsection we will quantize the model both in the gauge invariant
and hybrid variables and prove the equivalence between the quantum theories in
the two representations.
\subsection{Quantization}
Since the model is completely integrable the quantization can be performed
exactly. Furthermore, since we have reduced the system to a maximal set of gauge
invariant variables, the quantization in the Shanmugadhasan representation is
straightforward and the reduced and the Dirac approach are equivalent
\cite{quantum}. We will discuss the reduced method for the maximal gauge
invariant representation and the Dirac method for the hybrid representation for
simplicity.\footnote{See also Ref.\ \cite{LMMP} where the quantization is
achieved by mapping the interacting classical equations into sets of free
equations and then introducing the corrisponding intertwining operators.}

Let us consider first the gauge invariant canonical representation.  Since $T$
transforms linearly under gauge transformations, a natural gauge fixing is
\be
T=t-t_0\,.\label{gfI}
\ee
By imposing Eq.\ (\ref{gfI}) the Lagrange multiplier is
\be
\mu=1\,,
\ee
so we are working in the gauge proper time. The effective action is
\be
S_{\rm eff}=\int dt \left[\dot U V+\dot X W + \dot Y Z - H_{\rm eff}\right]\,,
\ee
where
\be
H_{\rm eff}=-H=0\,.
\ee
Let us quantize the model. In order to do this we must choose a representation
for the quantum observables. Taking into account Eqs.\ (\ref{obsII}) we choose
the representation in which  $\hat V$, $\hat W$ and $\hat Z$ are differential
operators with eigenvalues $v$, $w$ and $z$, respectively. The Hilbert space
measure is
\be
d[\varpi]=dudxdy\,,\label{measureII}
\ee
and $\hat V$, $\hat X$ and $\hat Z$ read
\be
\hat V=-i{\partial\over\partial u}\,,\qquad
\hat W=-i{\partial\over\partial x}\,,\qquad
\hat Z=-i{\partial\over\partial y}\,.\label{opII}
\ee
The Schr\"odinger equation reads
\be
H_{\rm eff}\Psi(u,x,y;t)=i{\partial~\over\partial t}\Psi(u,x,y;t)\,.\label{SchroI}
\ee
Since the effective Hamiltonian of the system is vanishing identically the wave
functions do not depend on $t$, i.e., $\Psi(u,x,y;t)=\Psi(u,x,y)$. (This
equation is nothing else that the Wheeler-De Witt (WDW) equation of the
system.) An orthonormal basis in the Hilbert space is given by the set of
eigenfunctions of the observables (\ref{opII}), namely
\be
\Psi(u,x,y)={1\over (2\pi)^{3/2}}\,e^{i(vu+wx+zy)}\,.
\label{waveII}
\ee
Let us briefly discuss the physical interpretation of Eq.\ (\ref{waveII}). From
the definition of $W$ in Eqs.\ (\ref{obsII}) we find $w=p_\beta$. Therefore,
the eigenstates (\ref{waveII}) with $w>0$ ($w<0$) corresponds to a growing
(decreasing) scale factor of the internal space. Analogously, $z>0$ ($z<0$)
corresponds to growing (decreasing) axion field. From Eqs.\ (\ref{fieldsII}) we
can define the operator corresponding to $\kappa^2$
\be
\hat\kappa^2={1\over 12}\left(-4{\partial^2~\over\partial u^2}+
{\partial^2~\over\partial x^2}\right)\,.
\label{kappaop}
\ee
The eigenvalues of the operators $\hat\kappa^2$ and of the operators in Eq.\
(\ref{opII}) are related by the relation $12\kappa^2=4v^2-w^2$. [See also Eq.\
(\ref{hypI}).] Using the previous results it is straightforward to identify the
wave functions that correspond to the classical cases {\it i)-iii)}.

Now let us turn to the hybrid canonical chart. From the constraint
(\ref{constraint-hybII}) it is natural to choose 
\be
d[\varpi]=dadbdcd\sigma\label{measure-hybII}
\ee
as measure in the wave function space.
Using this measure the operators $\hat p_a$, $\hat p_b$, $\hat p_c$ and $\hat
p_\sigma$ are
\be
\hat p_a=-i{\partial\over\partial a}\,,\quad
\hat p_b=-i{\partial\over\partial b}\,,\quad
\hat p_c=-i{\partial\over\partial c}\,,\quad
\hat p_\sigma=-i{\partial\over\partial\sigma}\,.
\label{op-hybII}
\ee
The WDW equation is
\be
\left[-{\partial^2~\over\partial a^2}
+{\partial^2~\over\partial b^2}
-{\partial^2~\over\partial c^2}
-e^{-2a}{\partial^2~\over\partial\sigma^2}\right]
\Psi(a,b,c,\sigma)=0\,.
\label{WDWII}
\ee
The WDW equation can be completely solved by the technique of separation of
variables. The general (bounded) solution is the superposition of wave
functions
$$
\Psi(a,b,c,\sigma)=\int dk_b dk_c dk_\sigma A(k_b,k_c,k_\sigma)
\psi(k_b,k_c,k_\sigma;a,b,c,\sigma)\,,
$$
\be
\psi(k_b,k_c,k_\sigma;a,b,c,\sigma)=N e^{\pm ib k_b}e^{\pm ic k_c}e^{\pm i\sigma k_\sigma}
K_{i\nu}(k_\sigma e^{-a})\,,\quad\nu=\sqrt{k_b^2-k_c^2}\,,\quad k_i\ge0\,,
\label{WDW-solII}
\ee
where $K_{i\nu}$ is the modified Bessel function of imaginary index $i\nu$
\cite{GR}. By properly choosing the normalization factor $N$, and fixing the
gauge using the $b$ degree of freedom, the eigenstates of the physical
Hamiltonian with energy $E=k^2_b$ read
\be
\psi_{k_b,k_c,k_\sigma}=\sqrt{\nu\sinh\pi\nu\over 2\pi^4}
e^{\pm ic k_c}e^{\pm i\sigma k_\sigma} K_{i\nu}(k_\sigma e^{-a})\,.
\label{waves-hybII}
\ee
Note that $\nu=\kappa$ on shell.
The wave functions (\ref{waves-hybII}) form an orthonormal basis in the Hilbert
space with (gauge-fixed) measure $d[\varpi]=dadcd\sigma$
\be
(\psi_1,\psi_2)=\int dadcd\sigma\, \psi_1^\star\psi_2=\delta(\nu_{1}-\nu_{2})
\delta(k_{c1}-k_{c2})\delta(k_{\sigma 1}-k_{\sigma 2})\,.
\ee
This completes the quantization in the hybrid representation.

Using the generating function (\ref{gfunctionII}) the equivalence between
the two sets of (gauge off-shell) solutions of the WDW equation (\ref{waveII})
and (\ref{WDW-solII}) can be proven. The relation between the wave functions in
the two representations is
\be
\Psi(a,b,c,\sigma)=\int d[\varpi(u,x,y;t)] e^{iF_1(a,b,c,\sigma;u,x,y,t)}
\Psi(u,x,y;t)\,,
\label{equivalence-waves}
\ee
where $d[\varpi(u,x,y;t)]$ is the measure in the space of wave functions. From
Eq.\ (\ref{measureII}) we know that $d[\varpi(u,x,y;t)]$ must be
\be
d[\varpi(u,x,y;t)]=dudxdydt\, t^{-p-2}\,.
\label{unknown-measureII}
\ee
The unknown parameter $p$ in Eq.\ (\ref{unknown-measureII}) represents possible
factor ordering ambiguities in the relation between the (gauge off-shell) wave
functions in the two representations. This ambiguity can be fixed by requiring
the equivalence of the (gauge on-shell) wave functions in the two
representations, thus proving the physical equivalence of the representations.
By substituting Eq.\ (\ref{waveII}) and Eq.\ (\ref{gfunctionII}) in Eq.\
(\ref{equivalence-waves}) and choosing $p=-1/2$ in Eq.\
(\ref{unknown-measureII}), after some algebra (details are given in Appendix B)
one obtains Eq.\ (\ref{WDW-solII}).

Let us briefly discuss the correspondence between the hybrid wave functions and
the classical solutions. The oscillating regions of the wave functions
correspond to the classically allowed regions of the configuration space. Along
$c$ and $\sigma$ directions the wave functions (\ref{waves-hybII}) are
described by plane waves. Along the $a$ direction the wave functions are
oscillating in the region
\be
0< e^{-a}\laq {\nu\over k_\sigma}\,.
\ee
This corresponds to the classically allowed region for the hybrid variable
$a$. (We have chosen $k_\sigma>0$ for simplicity.) Indeed, from the solutions
of the equations of motion we have
\be
0<e^{-a}={\kappa\over p_\sigma}[\cosh(\kappa\tau)]^{-1}\le
{\kappa\over p_\sigma}\,.
\ee
Finally, the wave functions go like $e^{\pm ia\nu}$ for large values of $a$.
With aid of Eqs.\ (\ref{fieldsII}), Eqs.\ (\ref{hybridII}), and Eqs.\
(\ref{gauge-to-hybII}), we find the relation between the quantum numbers $k_i$
and the classical parameters that characterize the behaviour of the classical
solution
\be
k_b=-2\sqrt{3}\xi\,,\qquad\qquad k_c={p_\beta\over 2\sqrt{3}}\,,\qquad\qquad\nu=\kappa\,.
\ee
Using the previous relations it is straightforward to identify the hybrid wave
functions that correspond to the classical cases {\it i)-iii)}. Finally,
starting from Eq.\ (\ref{waves-hybII}) quantum M-theory cosmology can be
investigated with aid of usual elementary quantum mechanics techniques. For
instance, wave packets can be constructed along the lines of Refs.\ \cite{WH}
and specific boundary conditions for the wave function of the universe can be
selected.

It is curious to note that the hybrid wave functions (\ref{waves-hybII}) are
formally equivalent to wormhole wave functions. [See e.g.\ \cite{WH} and
references therein.] It would be interesting to investigate whether this
analogy is purely accidental or conceals a deeper relation between M-theory
cosmology and wormhole physics.

\section{Invariant manifold II\label{caseII}}
In this case the phase space is six-dimensional and the canonical action is
\be
S_I=\int dt \left[\dot\alpha p_\alpha+\dot\phi p_\phi+\dot\beta
p_\beta-{\cal H}\right]\,,
\label{e1}
\ee
where the Hamiltonian is
\be
{\cal H}=\mu H\,,\qquad
H={1\over 24}\left(2p^2_\alpha-6p^2_\phi+p^2_\beta+
12Q^2e^{3\alpha-\phi-6\beta}\right)\,.\label{HI}
\label{e2}
\ee
The canonical equations of motion are
\be
\begin{array}{lll}
\displaystyle
\dot\alpha={p_\alpha\over 6}\,,\quad
&\displaystyle
\dot\phi=-{p_\phi\over 2}\,,\quad
&\displaystyle
\dot\beta={p_\beta\over 12}\,,\\\\
\displaystyle
\dot p_\alpha=-3{Q^2\over 2}e^{3\alpha-\phi-6\beta}\,,\quad
&\displaystyle
\dot p_\phi={Q^2\over 2}e^{3\alpha-\phi-6\beta}\,,\quad
&\displaystyle
\dot p_\beta=3Q^2e^{3\alpha-\phi-6\beta}\,.
\end{array}\label{eqsI}
\ee
The previous equations are supplemented by the constraint
\be
2p_\alpha^2-6p^2_\phi+p^2_\beta+12Q^2e^{3\alpha-\phi-6\beta}=0\,.
\label{constraintI}
\ee
Let us first discuss the case $Q\not=0$. The off-shell solution of the
canonical equations is 
\be
\begin{array}{lll}
\alpha&=&\displaystyle \alpha_0-{1\over 4}\ln
\left[\cosh\left(\kappa(\tau-\tau_0)\right)\right]
-\xi(\tau-\tau_0)\,,\\\\
p_\alpha&=&\displaystyle -{3\kappa\over 2}\tanh\left[\kappa(\tau-\tau_0)\right]
-6\xi\,,\\\\\
\phi&=&\displaystyle \phi_0-{1\over 4}\ln
\left[\cosh\left(\kappa(\tau-\tau_0)\right)\right]-
\chi(\tau-\tau_0)\,,\\\\
p_\phi&=&\displaystyle {\kappa\over 2}\tanh\left[\kappa(\tau-\tau_0)\right]
+2\chi\,,\\\\
\beta&=&\displaystyle \beta_0+{1\over 4}\ln
\left[\cosh\left(\kappa(\tau-\tau_0)\right)\right]-
\rho(\tau-\tau_0)\,,\\\\
p_\beta&=&\displaystyle 3\kappa\tanh\left[\kappa(\tau-\tau_0)\right]-12\rho\,,
\end{array}\label{solI}
\ee
where the constants of motion are related by
\be
\kappa^2+6\xi^2-2\chi^2+12\rho^2=2H\,,\qquad\kappa\not=0\,,
\label{hypII}
\ee
and (we choose $\kappa>0$ for simplicity)
\be
3\alpha_0-\phi_0-6\beta_0=2\ln\left({\kappa\over |Q|}\right)\,,\qquad
3\xi-6\rho-\chi=0\,.
\ee
On the (physical) shell $\chi=0$ is a degenerate trivial configuration of the
system because it implies $\kappa=0$, $\xi=0$, and $\rho=0$.

Analogously to the invariant manifold I, we can find a maximal set of gauge
invariant observables. In this case we expect four physical gauge invariant
observables because the system possesses three canonical degrees of freedom. A
possible choice is
\be
\displaystyle
\begin{array}{l}
\displaystyle
U=\alpha-\phi-{p_\alpha+3p_\phi\over 6\kappa}\arccosh\left({\kappa\over
|Q|}e^{-(3\alpha-6\beta-\phi)/2}\right)\,,\\\\
\displaystyle
V=-{1\over 2}\left(p_\alpha+3p_\phi\right)\,,\\\\
\displaystyle
X={1\over 4}(3\alpha+2\beta-\phi)-{3p_\alpha+p_\beta+3p_\phi\over 24\kappa}
\arccosh\left({\kappa\over |Q|}e^{-(3\alpha-6\beta-\phi)/2}\right)\,,\\\\
\displaystyle
W={1\over 2}\left(3p_\alpha+p_\beta+3p_\phi\right)\,,
\end{array}
\label{obsI}
\ee
where
\be
\kappa=\left[{1\over 16}\left(-p_\alpha+p_\beta-p_\phi\right)^2+
Q^2e^{3\alpha-6\beta-\phi}\right]^{1/2}\,.
\ee
The quantities (\ref{obsI}) have been chosen such that they satisfy
the canonical Poisson brackets
\be
\begin{array}{lll}
\left[U,V\right]_P=1\,,&&\left[X,W\right]_P=1\,.
\end{array}\label{canonicalPBI}
\ee
The conjugate of the Hamiltonian that completes the Shanmugadhasan chart is
\be
T={1\over\kappa}
\arccosh\left({\kappa\over |Q|}
e^{-(3\alpha-6\beta-\phi)/2}\right)\,.\label{TI}
\ee
Using Eqs.\ (\ref{HI}), (\ref{obsI}) and (\ref{TI}) we find the relation
between the gauge invariant observables and the constants of motion in Eqs.\
(\ref{solI})
\be
\begin{array}{lll}
\displaystyle \alpha_0={1\over 2}(3X-U)+{1\over 4}
\ln{\kappa\over |Q|}\,,
&\displaystyle 
\beta_0={X\over 2}-{1\over 4}
\ln{\kappa\over |Q|}\,,
&\displaystyle 
\phi_0={3\over 2}(X-U)+{1\over 4}
\ln{\kappa\over |Q|}\,,\\\\
\displaystyle
\xi=-{1\over 6}(V+3W/4)\,,
&\displaystyle
\rho=-{W\over 24}\,,
&\displaystyle
\chi=-{1\over 2}(V+W/4)\,,
\end{array}\label{fieldsI}
\ee
and
\be
\kappa=\left[{1\over 12}(4V^2-W^2)+2H\right]^{1/2}\,.
\ee

The relation between the invariant submanifold I and the invariant submanifold
II can be made manifest in the hybrid canonical chart. For the invariant
submanifold II the hybrid variables $(a,b,c)$ are 
\be
\begin{array}{lll}
\displaystyle
a=-{1\over 2}\left(3\alpha-6\beta-\phi\right)\,,
&\displaystyle
b=\sqrt{3}(\alpha-\phi)\,,
&\displaystyle
c={\sqrt{3}\over 2}\left(3\alpha+2\beta-\phi\right)\,,\\\\
\displaystyle
p_a={1\over 4}\left(-p_\alpha+p_\beta-p_\phi\right)\,,
&\displaystyle
p_b=-{1\over 2\sqrt{3}}\left(p_\alpha+3p_\phi\right)\,,
&\displaystyle
p_c={1\over 4\sqrt{3}}\left(3p_\alpha+p_\beta+3p_\phi\right)\,.
\end{array}\label{hybridI}
\ee
Indeed, using the hybrid variables the constraint (\ref{constraintI}) reads
(we have divided by a factor $12$)
\be
p_a^2-p^2_b+p^2_c+Q^2 e^{-2a}=0\,.
\label{constraint-hybI}
\ee
Moreover, the gauge invariant canonical variables (\ref{obsI}) are related to
the hybrid variables (\ref{hybridI}) by the canonical transformation 
\be
\displaystyle
\begin{array}{ll}
\displaystyle
U={1\over\sqrt{3}}\left[b+{p_b\over\kappa}
\arccosh\left({\kappa\over |Q|}e^{a}\right)\right]\,,
&\displaystyle
V=\sqrt{3}p_b\,,\\\\
\displaystyle
X={1\over 2\sqrt{3}}\left[c-{p_c\over\kappa}
\arccosh\left({\kappa\over |Q|} e^{a}\right)\right]\,,
&\displaystyle
W=2\sqrt{3}p_c\,,\\\\
\displaystyle
T={1\over\kappa}
\arccosh\left({\kappa\over |Q|} e^{a}\right)\,,
&\displaystyle
H={1\over 2}\left(p_a^2-p^2_b+p^2_c+Q^2 e^{-2a}\right)\,,
\end{array}\label{gauge-to-hybI}
\ee
where $\kappa=\sqrt{p_a^2+Q^2e^{-2a}}$. Note that this transformation coincides
with the section $p_\sigma=Q$ of the transformation (\ref{gauge-to-hybII}).

\begin{figure}
\centerline{\epsfxsize=4.0in \epsffile{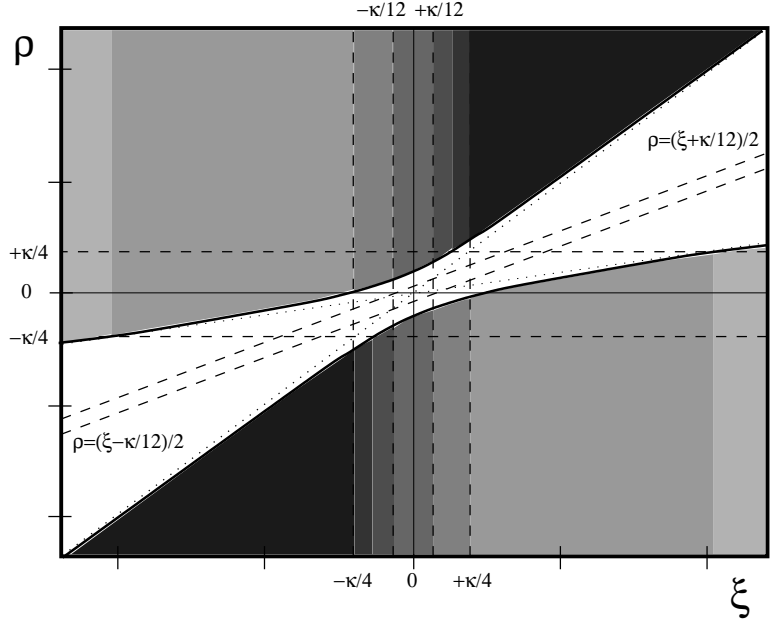}}
\caption{\small Parameter space for invariant submanifold II. The physical
points are represented by the two branches of the hyperbola $\rho=\left(3\xi\pm
\sqrt{4\xi^2+5\kappa^2/12}\right)/5$.The colored regions determine on the two
branches of the hyperbola the different physical cases described in the text.}
\label{figII}
\end{figure}

The qualitative discussion of the physical properties of the solution proceeds
similarly to that of the first invariant manifold.\footnote{See also Ref.\
\cite{KKO} where eleven-dimensional cosmological solutions with and without
four-form charges are investigated. Note that Kaloper, Kogan and Olive work in
the eleven-dimensional supergravity frame whereas we work in the string frame.
The relation between the quantities used in our paper and those used in Ref.\
\cite{KKO} (denoted with a twiddle) are $\alpha=\tilde\alpha+\tilde\gamma/2$,
$\beta=\tilde\beta+\tilde\gamma/2$, $\Phi_{10}=3\tilde\gamma$, and $N=\tilde n
e^{\tilde\gamma/2}$.} The dynamics of the model is determined by $\kappa$ and
$\xi$. At fixed $\kappa$, the physical points of the model are determined in
the plane ($\xi,\rho$) by the two branches of the hyperbola (\ref{hypII})
$$
\hbox{Upper branch: } \rho={3\over 5}\xi+{1\over 5}
\sqrt{4\xi^2+{5\kappa^2\over 12}}\,;\qquad
\hbox{Lower branch: } \rho={3\over 5}\xi-{1\over 5}
\sqrt{4\xi^2+{5\kappa^2\over 12}}\,.
$$
The upper branch is characterized by an ever increasing four-dimensional
effective coupling $g$, which evolves from zero at $\tau\to-\infty$ to infinity
at $\tau\to\infty$. Conversely, the lower branch is characterized by an ever
decreasing four-dimensional effective coupling $g$, which starts with an
infinite value at $\tau\to-\infty$ and vanishes at $\tau\to\infty$. For each of
the two branches we distinguish three different dynamical behaviours of the
external geometry according to the value of $\xi$ (see Fig.\ \ref{figII}).
For the upper branch we have:
\begin{itemize}
\item[{\it i)}] Ever expanding external spacetime. This happens for
$\xi\le-\kappa/4$. The ten-dimensional coupling constant is zero at
$\tau\to-\infty$ and becomes infinite at $\tau\to\infty$. The Hubble constant
is always positive and vanishes at $\tau=-\infty$ where the external scale
factor is also zero. At large times ($\tau\to\infty$) both the Hubble constant
and the external scale factor become infinite except for the limiting value
$\xi=-\kappa/4$ for which they reach zero and a finite nonzero value,
respectively. The radius of the internal space is infinite at large times and
zero, finite, and infinite for $\xi<-\kappa(3/4+1/\sqrt{3})$,
$\xi=-\kappa(3/4+1/\sqrt{3})$, and $-\kappa(3/4+1/\sqrt{3})<\xi\le-\kappa/4$ at
$\tau\to-\infty$, respectively.
\item[{\it ii)}] Ever contracting external spacetime. This happens for
$\xi\ge\kappa/4$. The ten-dimensional coupling constant is infinite at
$\tau\to-\infty$ and vanishes at $\tau\to\infty$. The Hubble constant is always
negative, in particular is zero at $\tau=-\infty$ and infinite at
$\tau=\infty$ where the external scale factor is zero. At small times
($\tau\to-\infty$) the external scale factor is infinite except for the
limiting value $\xi=\kappa/4$ for which is finite nonzero. The
radius of the internal space is infinite at $\tau=-\infty$ and zero at
$\tau=\infty$. 
\item[{\it iii)}] Bouncing external spacetime ($-\kappa/4<\xi<\kappa/4$). In
this case the radius of the external spacetime is zero at $\tau\to-\infty$,
reaches a maximum, and then decreases to zero at $\tau\to\infty$. The Hubble
constant evolves from zero to $-\infty$ passing through a positive maximum. The
ten-dimensional coupling constant and the scale factor of the internal space
evolve as follows
\begin{itemize}
\item[a)] For $-\kappa/4<\xi\le-\kappa/12$ $g_{10}$ evolves from zero [from a
finite nonzero value for the limiting value $\xi=-\kappa/12$] to infinity and
$R_{T^6}$ bounces from infinity to infinity;
\item[b)] For $-\kappa/12<\xi\le\kappa/12$ both $g_{10}$ and $R_{T^6}$ bounce
from infinity to infinity [$g_{10}$ to a finite nonzero value for the limiting
value $\xi=\kappa/12$] (see Fig.\ \ref{figIIab});
\begin{figure}
\centerline{\epsfysize=150pt \epsffile{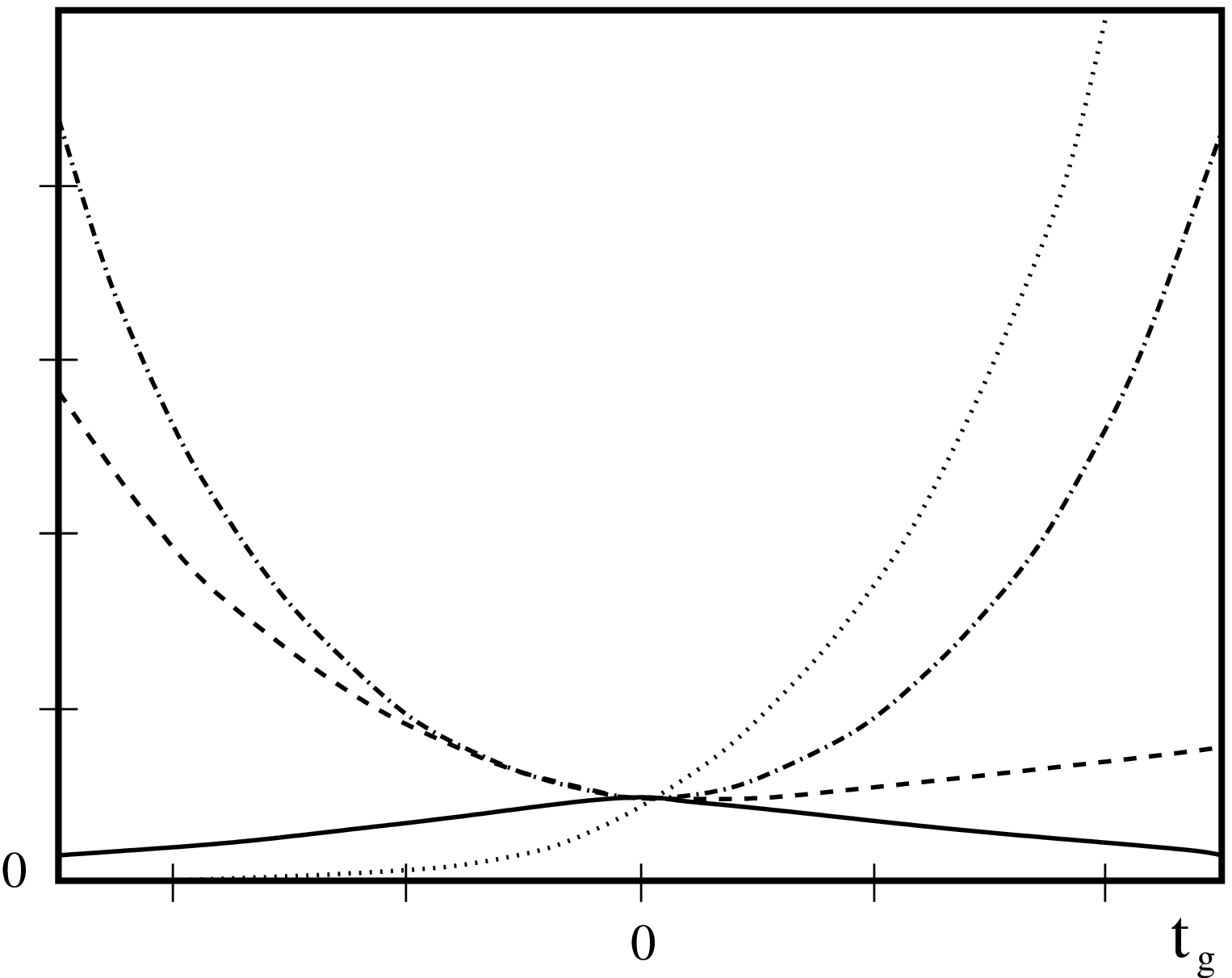}\qquad
\epsfysize=150pt \epsffile{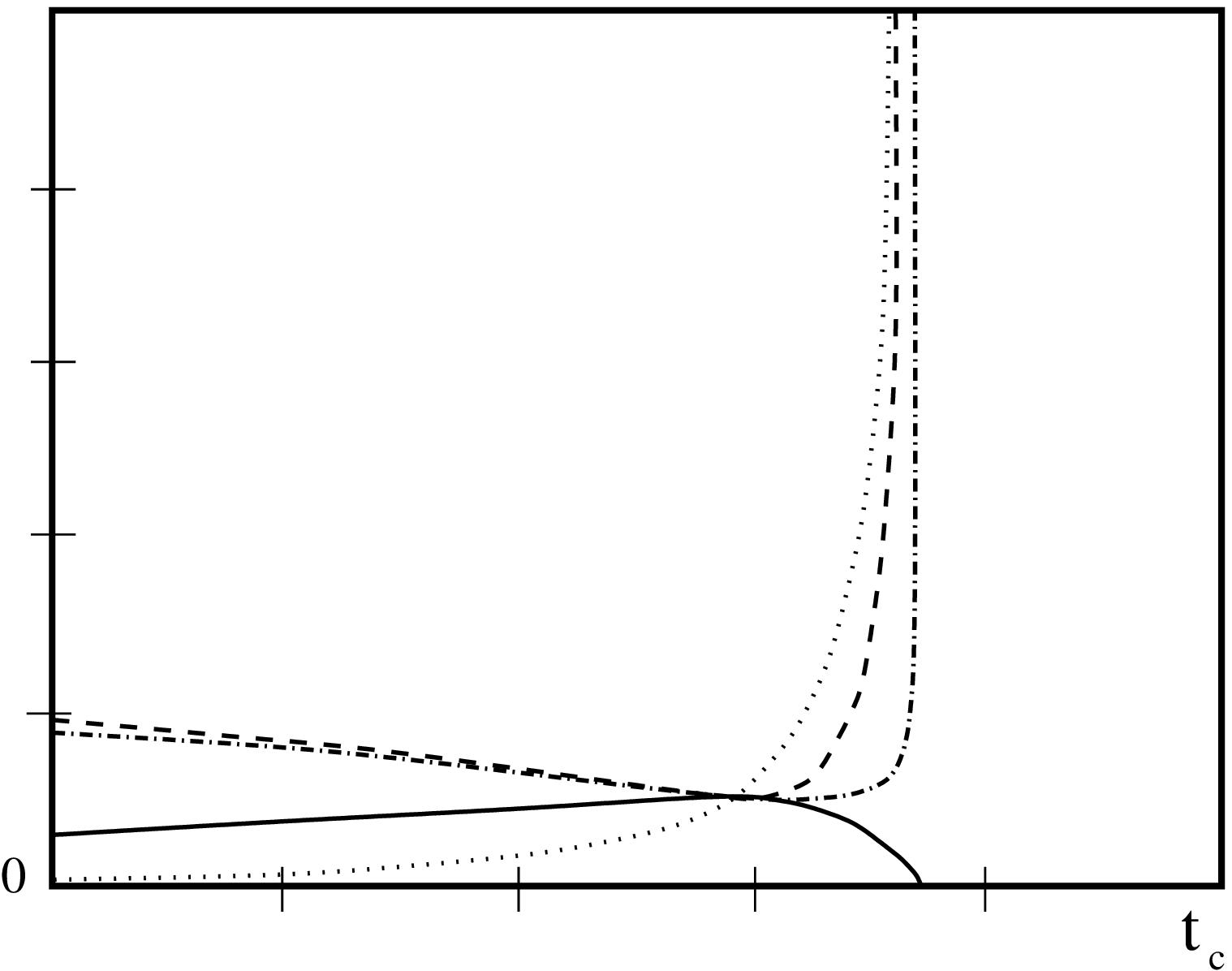}}
\caption{\small Graphical representation of the dynamics of the invariant
submanifold II for $-\kappa/12<\xi<\kappa/12$ (upper branch) as function of
the gauge proper time $t_g$ and of the cosmic proper time $t_c$. The external
scale factor is vanishing at small times, reaches a maximum, and then decreases
to zero at large times. Both the  internal scale factor and the coupling
constant $g_{10}$ are infinite at small times, decrease to a minimum value, and
then increase again to an infinite value for large times.
(See Fig.\ \ref{figIab} for the key to the plotted lines.)}
\label{figIIab}
\end{figure}
\item[c)] For $\kappa/12<\xi<\kappa(3/4-1/\sqrt{3})$ $g_{10}$ evolves from
infinity to zero and $R_{T^6}$ bounces from infinity to infinity;
\item[d)] For $\kappa(3/4-1/\sqrt{3})\le\xi<\kappa/4$ both $g_{10}$ and
$R_{T^6}$ evolve from infinity to zero [$R_{T^6}$ to a finite nonzero value for
the limiting value $\xi=\kappa(3/4-1/\sqrt{3})$].
\end{itemize}
\end{itemize}
For the lower branch we have:
\begin{itemize}
\item[{\it i)}] Ever expanding external spacetime for $\xi\le-\kappa/4$. The
ten-dimensional coupling constant is zero at $\tau\to-\infty$ and becomes
infinite at $\tau\to\infty$. The Hubble constant is always positive and
vanishes at $\tau=\infty$ where the external scale factor is infinite except
for the limiting value $\xi=-\kappa/4$ for which it reaches a finite nonzero
value. At small times ($\tau\to-\infty$) the Hubble constant is infinite and
the external scale factor is zero. The radius of the internal space evolves from
zero to infinity (see Fig.\ \ref{figIIcd}).
\begin{figure}
\centerline{\epsfysize=150pt \epsffile{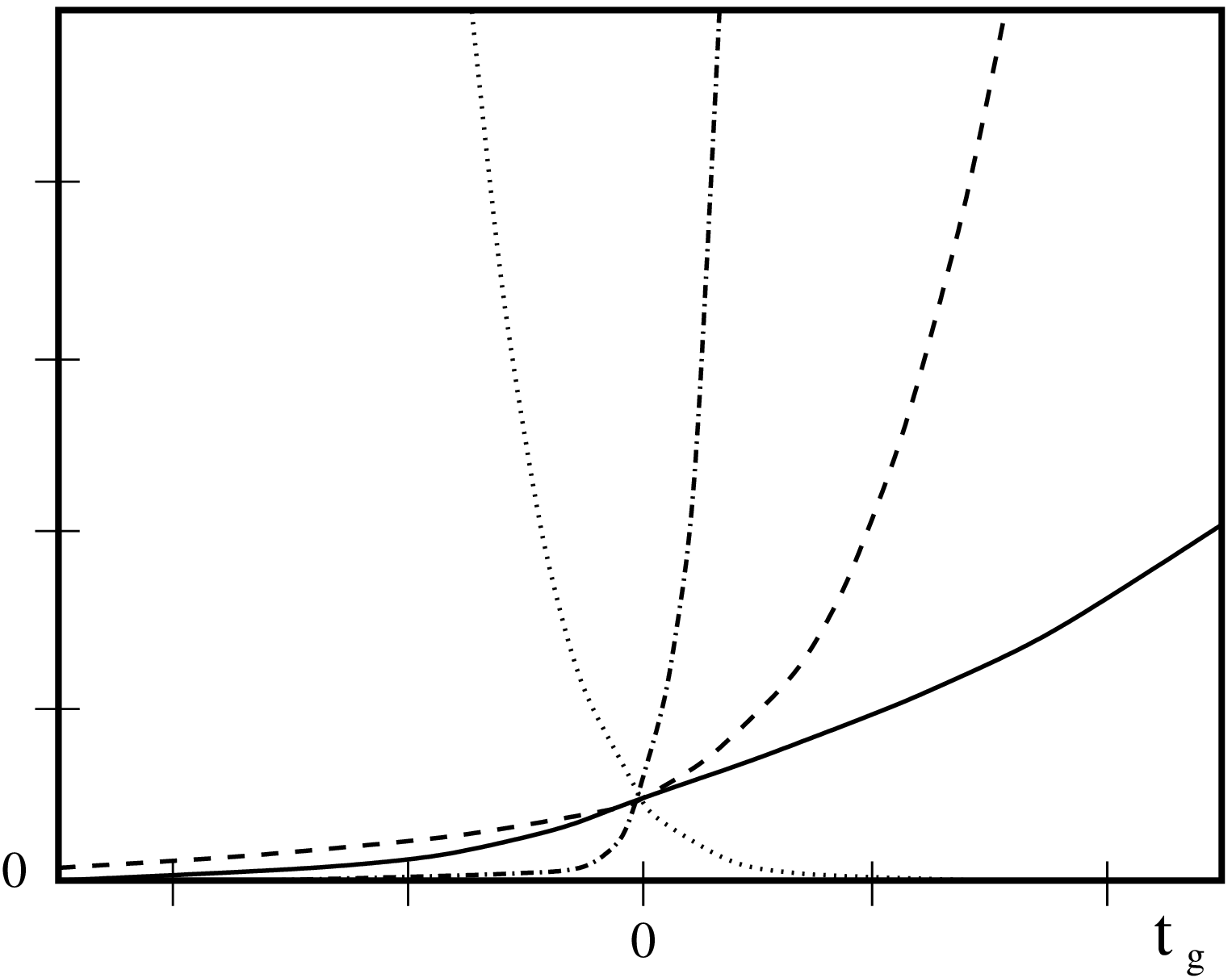}\qquad
\epsfysize=150pt \epsffile{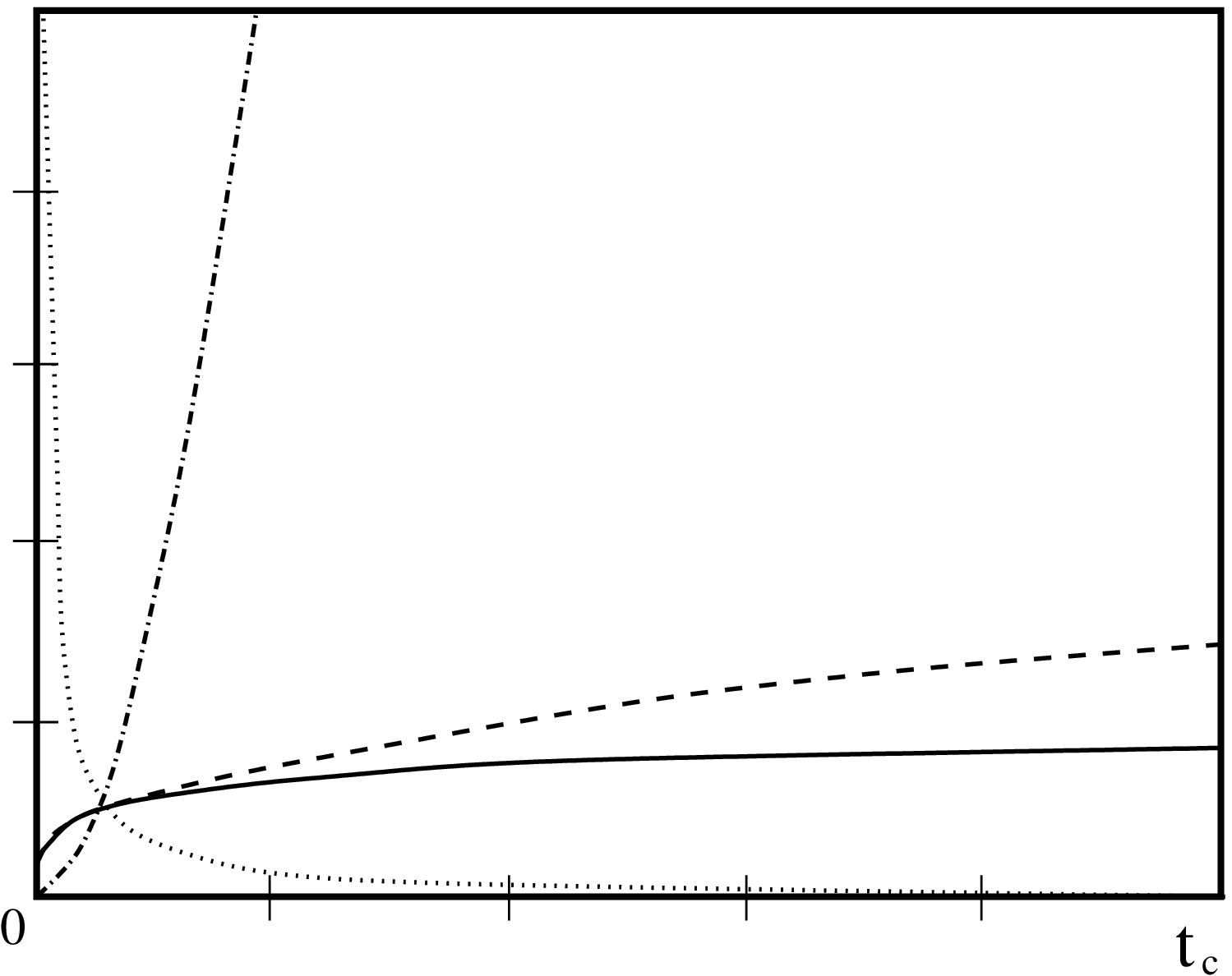}}
\caption{\small Graphical representation of the dynamics of the invariant
submanifold II for $\xi<-\kappa/4$ (lower branch) as function of the gauge
proper time $t_g$ and of the cosmic proper time $t_c$. The scale factor of the
external spacetime, the scale factor of the internal space and the
ten-dimensional coupling constant $g_{10}$ are ever increasing. The coupling
constant $g$ is ever decreasing. (See Fig.\ \ref{figIab} for the key to the
plotted lines.)}
\label{figIIcd}
\end{figure}
\item[{\it ii)}] Ever contracting external spacetime for $\xi\ge\kappa/4$. The
ten-dimensional coupling constant is infinite at $\tau\to-\infty$ and vanishes
at $\tau\to\infty$. Both the external scale factor and the Hubble constant
(always negative) vanish at large times.  At small times the radius of the
external spacetime is infinite and $H(-\infty)=-\infty$ except for the limiting
value $\xi=\kappa/4$ for which the radius of the external spacetime and $H$ are
finite nonzero and zero, respectively. In this case $H$ evolves through a
negative minimum before going to zero at large times. The radius of the
internal space is infinite for $\tau=-\infty$ and infinite, finite nonzero, and
zero at $\tau=\infty$ for $\kappa/4<\xi<\kappa(3/4+1/\sqrt{3})$,
$\xi=\kappa(3/4+1/\sqrt{3})$, and $\xi>\kappa(3/4+1/\sqrt{3})$, respectively.
\item[{\it iii)}] Bouncing external spacetime for $-\kappa/4<\xi<\kappa/4$. In
this case the radius of the external spacetime is zero at $\tau\to-\infty$,
reaches a maximum, and then decreases to a zero value at $\tau\to\infty$. The
Hubble constant evolves from infinite to zero through a negative minimum. The
ten-dimensional coupling constant and the scale factor of the internal space
evolve as follows
\begin{itemize}
\item[a)] For $-\kappa/4<\xi\le-\kappa(3/4-1/\sqrt{3})$ $g_{10}$ evolves from
zero to infinity and $R_{T^6}$ evolves from zero [from a finite nonzero value
for the limiting value $\xi=-\kappa(3/4-1/\sqrt{3})$] to infinity;
\item[b)] For $-\kappa(3/4-1/\sqrt{3})<\xi\le-\kappa/12$ $g_{10}$ evolves from
zero [from a finite nonzero value for the limiting value $\xi=-\kappa/12$] to
infinity and $R_{T^6}$ bounces from infinity to infinity;
\item[c)] For $-\kappa/12<\xi\le\kappa/12$ both $g_{10}$ and $R_{T^6}$ bounce
from infinity to infinity [$g_{10}$ to a finite nonzero value for
$\xi=\kappa/12$];
\item[d)] For $\kappa/12<\xi<\kappa/4$ $g_{10}$ evolves from infinity to zero
and $R_{T^6}$ bounces from infinity to infinity.
\end{itemize}
\end{itemize}
From the physical point of view cases {\it i)} (ever expanding external
spacetime) are not viable because both the eleven dimension and the internal
space always decompactify at large times. Bouncing external spacetimes can
instead give a more realistic description of our universe. In particular, case
{\it iii)}-d) of the upper branch is characterized by an ever decreasing
ten-dimensional coupling and an ever decreasing internal scale factor. This
property guarantees that a large-size external spacetime is weak-coupled and
its internal space is compactified. Moreover, a bouncing external
spacetime is always characterized by early accelerated expansion. Models
with bouncing $g_{10}$ and $R_{T^6}$ are probably also viable from the physical
point of view, albeit they might require some kind of fine tuning of the free
parameters to actually describe a realistic universe.
\subsection{Quantization}
The quantization of the system proceeds analogously to the case I. (See also
the footnote in Subsection 3.2.) Let us start again with the Shanmugadhasan
representation. By imposing the gauge fixing (\ref{gfI}) we obtain the
effective action
\be
S_{\rm eff}=\int dt \left[\dot U V+\dot X W - H_{\rm eff}\right]\,,
\ee
with vanishing effective Hamiltonian $H_{\rm eff}=-H$. We work in the
representation in which $\hat V$, $\hat W$ are differential operators with
eigenvalues $v$ and $w$, respectively. The Hilbert space measure is
\be
d[\varpi]=dudx\,,\label{measureI}
\ee
and $\hat V$, $\hat W$ are
\be
\hat V=-i{\partial\over\partial u}\,,\qquad
\hat W=-i{\partial\over\partial x}\,,\label{opI}
\ee
An orthonormal basis in the Hilbert space is given by the set of eigenfunctions
of the observables (\ref{opI}), namely
\be
\Psi(u,x,y)={1\over 2\pi}\,e^{i(vu+wx)}\label{waveI}
\ee

In the hybrid representation the quantization is formally identical to that of
case I. So we give here only the result. The Hamiltonian eigenstates with
energy $E=k_b^2$ are
\be
\psi(k_b,k_c;a,c)=\sqrt{\nu\sinh\pi\nu\over 2\pi^4}
e^{\pm ic k_c}
K_{i\nu}(Q e^{-a})\,,\quad\nu=\sqrt{k_b^2-k_c^2}\,,\quad k_i\ge0\,,
\label{WDW-solI}
\ee
The set of wave functions (\ref{WDW-solI}) forms an orthonormal basis w.r.t.\
gauge-fixed Hilbert space with measure $d[\varpi]=dadc$. Starting from Eqs.\
(\ref{WDW-solI}) the quantum mechanics of the invariant submanifold II can be
investigated.

\begin{figure}
\centerline{\epsfxsize=4.0in
\epsffile{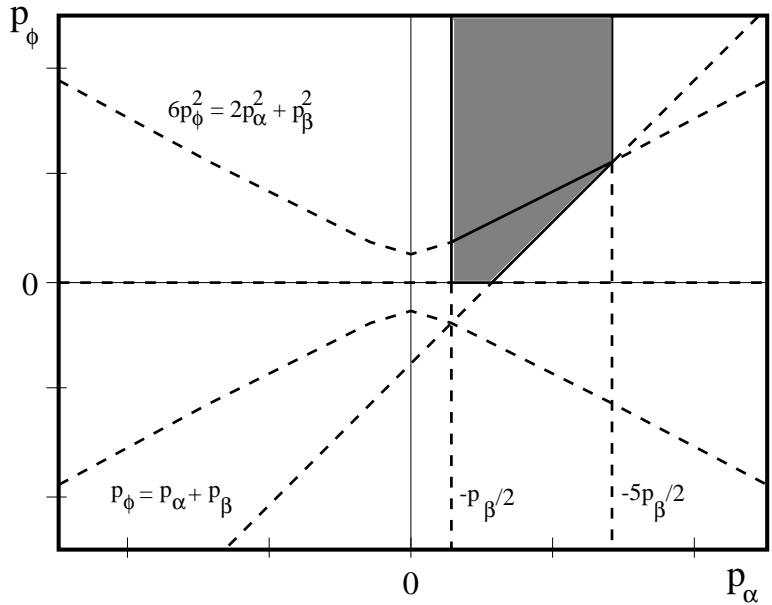}}
\caption{\small Parameter space for the degenerate case $S_{I\cap II}$. The
portion of the hyperbola in the colored region represents geometries with
contracting internal space ($p_\beta<0$) and expanding external spacetime which
evolve from a strong coupling regime ($g\to\infty$, $g_{10}\to\infty$) at
$\tau=-\infty$ to a weak coupling regime ($g\approx 0$, $g_{10}\approx 0$) at
$\tau=\infty$.}
\label{figIII}
\end{figure}

Finally, let us discuss the degenerate case $S_{I\cap II}$ ($Q=0$). The 
general (Kasner) solution is (see also \cite{KKO} and footnote 4)
\be
\alpha=\alpha_0+{p_\alpha\over 6}(\tau-\tau_0)\,,\qquad
\phi=\phi_0-{p_\phi\over 2}(\tau-\tau_0)\,,\qquad
\beta=\beta_0+{p_\beta\over 12}(\tau-\tau_0)\,,
\label{solIII}
\ee
where $p_\alpha$, $p_\phi$, and $p_\beta$ are constants related by the
condition
\be
2p_\alpha^2-6p^2_\phi+p^2_\beta=2H\,.
\label{constraintIII}
\ee
Note that Eqs.\ (\ref{solI}) reduce to Eqs.\ (\ref{solIII}) with
$\kappa=0$ and a redefinition of the integration constants. The on-shell
dynamics is identical to the dynamics of a Klein-Gordon particle moving in a
three-dimensional Minkowski space with timelike coordinate $\phi/\sqrt{6}$ and
spacelike coordinates $\alpha/\sqrt{2}$ and $\beta$. In this case the gauge and
hybrid canonical variables coincide and the quantization of the system is
straightforward. Classically, the $S_{I\cap II}$ model describes a spacetime
with expanding, contracting, and constant internal (external) space for
$p_\beta>0$ ($p_\alpha>0$), $p_\beta<0$ ($p_\alpha<0$), and $p_\beta=0$
($p_\alpha=0$), respectively. The sign of $p_\phi$ determines the strong and
weak coupling regions of the model. In particular, a geometry with contracting
internal space ($p_\beta<0$) and expanding external spacetime evolves from a
strong region to a weak region for $p_\phi>p_\alpha+p_\beta>0$ and
$-p_\beta/2<p_\alpha<-5p_\beta/2$ (see Fig.\ \ref{figIII}).  
\section{Conclusions\label{concl}}
In this paper we have analysed a spatially flat, four-dimensional cosmological
model derived from the M-theory effective action [Eq.\ (\ref{b1})]. The
eleven-dimensional metric is first compactified on a one-dimensional circle to
obtain the type IIA superstring effective action and then on a six-torus to
obtain the effective four-dimensional theory. In our investigation we
concentrated the attention on the boundary of the physical phase space of the
theory which is constituted by two invariant submanifolds, where either the
axion field (from the NS/NS sector) is negligible or the four-form R/R field
strength is irrelevant. In our discussion we have heavily employed the
canonical formalism. This approach makes the analysis of the features of the
classical solution extremely simple and allows a straightforward quantization
of the theory.

We found regions in the moduli space where a four-dimensional FRW universe
evolves from a strong coupling regime towards a weak coupling regime, both
internal six-volume and eleven-dimension contracting. In some cases, the
dynamics is characterized by an early accelerated (inflationary) expansion with
the spacetime eventually approaching a standard FRW decelerated expansion. Such
scenario happens when the pseudo-scalar axion field dominates and the R/R field
strength is negligible. (See also Ref.\ \cite{CLW}.) When the axion is
subdominant w.r.t. the R/R field, different cosmological scenarios appear on
the scene. In this case an ever expanding external spacetime is characterized
at large times by an infinite ten-dimensional coupling constant (the size of
eleven dimension). This seems to rule out this model from a physical point of
view. On the contrary, for bouncing external spacetimes, where the external
dimensions recollapse after initially expanding, geometries with early
accelerated expansion, decreasing ten-dimensional coupling constant and
decreasing internal space are found. 

The quantization of the two invariant submanifolds can be performed exactly and
the Hilbert space of states is derived. In the quantum framework our analysis
allows to identify the quantum states that correspond to the different
classical behaviours. In the hybrid representation we have identified regions
in the space of parameters where the wave function of the universe is either
oscillating or exponentially decaying. These regions are determined by  the
inverse exponential function of the four-dimensional (unshifted) dilaton, i.e.,
by the four-dimensional string coupling, and correspond to classically allowed
and classically forbidden regions, respectively. Starting from the Hilbert
space of states, the quantum mechanics of M-theory cosmology can be constructed
with aid of usual elementary quantum mechanics techniques.

The results summarized in the preceeding paragraphs strengthen and extend
previous results obtained within different approaches \cite{LMMP}-\cite{BCLN}.
Although the models investigated in this paper represent a severe, yet
consistent, truncation of the original eleven-dimensional action Eq.\
(\ref{b1}), new interesting physical output can be extracted. In particular,
the quantum mechanics of M-theory that has been presented here may constitute
the basis for further insights in high-energy (quantum) M-theory cosmology.
{\em If} M-theory is the ultimate theory of quantum gravity, M-theory quantum
cosmology is a subject which is certainly worth exploring. From this point of
view related topics of investigation which deserve attention are the study of
inhomogenous perturbations of the fields and a complete analysis of FRW models
in the full phase space of the theory, when all form fields are excited. In the
latter case, as Damour and Henneaux have recently observed \cite{DH}, the
dynamics may be qualitatively different and lead to potentially new theoretical
and observational effects.
\section*{Acknowledgments}
We are very grateful to M.\ Gasperini, M.\ Henneaux, N.\ Kaloper, J.\ Lidsey,
C.\ Ungarelli, A.\ Vilenkin and D.\ Wands for interesting discussions and
useful comments. This work is supported by grants ESO/PROJ/12\-58/98,
CERN/P/FIS/15190/1999, Sapiens-Proj32327/99. M.C.\ is partially supported by
the FCT grant Praxis XXI BPD/20166/99.
\section*{Appendix A. Comparison with notations of BCLN.}
In this Appendix we compare our notations with those of Ref.\ \cite{BCLN}. This
allows a straightforward interpretation of our results in the formalism of
BCLN.

Let us first consider the invariant submanifold I. Using Eqs.\ (\ref{eqsII}) and
(\ref{taub}), and Eq.\ (26) of Ref.\ \cite{BCLN} the BCLN variables $\xb$, $\yb$
and $\Ob$ read
\be
\xb=-{p_\alpha\over \sqrt{3}p_\phi}\,,\qquad 
\yb=-{p_\beta\over \sqrt{6}p_\phi}\,,\qquad
\Ob=2\left({p_\sigma\over p_\phi}\right)^2e^{-2(3\alpha+\phi)}\,,
\label{A1}
\ee
respectively. Moreover we have the important relation
\be
\zb\equiv 1-\xb^2-\yb^2-\Ob+{4\over p_\phi^2}H=0\,.\label{zbII}
\ee
Differentiating Eqs.\  (\ref{A1}) w.r.t.\ $\phi$ and recalling Eq.\
(\ref{taub}) we have (on shell)
\ba
{d\xb\over d\Tb}&=&\Ob(\xb+\sqrt{3})\,,\nonumber\\
{d\yb\over d\Tb}&=&\Ob\yb\,,\label{eqbII}\\
{d\Ob\over d\Tb}&=&-2\Ob(\xb^2+\yb^2+\sqrt{3}\xb)\,,\nonumber
\ea
where we have used Eq.\ (\ref{zbII}) and the equations of motion (\ref{eqsII}).
Eqs.\ (\ref{eqbII}) coincide with Eqs.\ (28)-(29) of Ref.\ \cite{BCLN} for the
invariant submanifold I. Since $\Ob\ge 0$ from Eq.\ (\ref{zbII}) the physical
region in the phase space is 
\be
\xb^2+\yb^2\le 1\,,
\ee
thus recovering the BCLN result. Finally, from (\ref{eqbII}) we obtain
\be
{d~~~~~~\over d\Tb}\left(\xb+\sqrt{2}\yb+\sqrt{3}\right)=\Ob
\left(\xb+\sqrt{2}\yb+\sqrt{3}\right)\,.\label{eqinvII}
\ee
Equation (\ref{eqinvII}) coincides with Eq.\ (33) of Ref.\ \cite{BCLN} for
$z=0$.

BLCN present the analitical solution for the invariant submanifold I (see Eq.\
(35) of Ref.\ \cite{BCLN}). Using Eq.\ (\ref{A1}) and the solutions of the
Eqs.\ of motion (\ref{solII}) it is straightforward to obtain Eq.\ (35) of
Ref.\ \cite{BCLN} where
\be
{y_0\over x_0+\sqrt{3}}={p_\beta\over 12\sqrt{2}\xi}\,.
\ee

Let us now consider the invariant submanifold II. Using Eqs.\ (\ref{eqsI}) and
(\ref{taub}), and Eq.\ (26) of Ref.\ \cite{BCLN} the BCLN variables $\xb$, $\yb$
and $\zb$ read
\be
\xb=-{p_\alpha\over \sqrt{3}p_\phi}\,,\qquad 
\yb=-{p_\beta\over \sqrt{6}p_\phi}\,,\qquad
\zb=2\left({Q\over p_\phi}\right)^2e^{3\alpha-\phi-6\beta}\,,\label{xyz}
\ee
respectively. Using Eqs.\ (\ref{xyz}) we find 
\be
\Ob\equiv 1-\xb^2-\yb^2-\zb+{4\over p_\phi^2}H=0\,,\label{ObI}
\ee
thus $\Ob$ is vanishing on-shell as expected for the invariant submanifold II.
Since $\zb\ge 0$ the physical region in the phase space is 
\be
\xb^2+\yb^2\le 1\,,
\ee
where the equality holds iff $Q=0$. 

Differentiating the previous equations w.r.t.\ $\phi$ and
recalling Eq.\ (\ref{taub}) we have
\ba
{d\xb\over d\Tb}&=&{\zb\over 2}\left(\xb-\sqrt{3}\right)\,,\nonumber\\
{d\yb\over d\Tb}&=&{\zb\over 2}\left(\yb+\sqrt{6}\right)\,,\label{eqbI}\\
{d\zb\over d\Tb}&=&\zb\left(\zb-1-\sqrt{6}\yb+\sqrt{3}\xb\right)\,.\nonumber
\ea
Eqs.\ (\ref{eqbI}) coincide with Eqs.\ (28)-(29) of Ref.\ \cite{BCLN} for the
invariant submanifold II. [Remember Eq.\ (\ref{ObI})]. Finally, we have
\be
\xb+\sqrt{2}\yb+\sqrt{3}={1\over\sqrt{3}p_\phi}(3p_\phi-p_\alpha-p_\beta)\,,
\label{inv}
\ee
and by differentiating Eq.\ (\ref{inv})
\be
{d~~~~~~\over d\Tb}\left(\xb+\sqrt{2}\yb+\sqrt{3}\right)={\zb\over 2}
\left(\xb+\sqrt{2}\yb+\sqrt{3}\right)\,.\label{eqinvI}
\ee
Equation (\ref{eqinvI}) coincides with Eq.\ (33) of Ref.\ \cite{BCLN} when
$\Ob=0$.

BLCN present also the analitical solution for the second invariant submanifold
(Eq.\ (34) of Ref.\ \cite{BCLN}). Again, it is straightforward to obtain the
BCLN result where
\be
{y_0+\sqrt{6}\over x_0-\sqrt{3}}=\sqrt{2}\;{\chi+\rho\over\xi-\chi}\,.
\ee
This completes the comparison to the BCLN variables. 
\section*{Appendix B. Proof of the equivalence between gauge invariant and hybrid
quantum representations.}
Let us consider Eq.\ (\ref{equivalence-waves}). By substituting Eq.\
(\ref{waveII}) and Eq.\ (\ref{gfunctionII}) the wave function in the
$(a,b,c,\sigma)$ representation can be written
\be
\Psi(a,b,c,\sigma)=N\int dt t^{-p-2} I_1(u,t;b) I_2(x,t;c) I_3(y,t;a,\sigma)
\label{B1}\,,
\ee
where
\ba
I_1(u,t;b)&=& \int du \exp\left[iuv+i{(\sqrt{3}u-b)^2\over 2t}\right]\,,\\
I_2(x,t;c)&=& \int dx \exp\left[ixw-i{(2\sqrt{3}x-c)^2\over 2t}\right]\,,\\
I_3(y,t;a,\sigma)&=& \int dy
\exp\left[iyz-i{\arccosh^2\sqrt{1+e^{2a}(y-\sigma)^2}\over 2t}\right]\,,\\
\ea
and $N$ is a normalization factor. The evaluation of the integrals $I_1$ and
$I_2$ is immediate. The result is\footnote{Here and throughout the Appendix
neglect overall constant factors that are reabsorbed in $N$.} 
\ba
I_1&=&\sqrt{t} \exp\left[i\left({bv\over\sqrt{3}}-{tv^2\over 6}\right)\right]\,,\\
I_2&=&\sqrt{t} \exp\left[i\left({cw\over 2\sqrt{3}}+{tw^2\over
24}\right)\right]\,.
\ea
Substituting the previous results in Eq.\ (\ref{B1}) and using the new variable
$\bar y=y-\sigma$, Eq.\ (\ref{B1}) is cast in the form
\be
\Psi(a,b,c,\sigma)=N\exp\left[i\left({bv\over\sqrt{3}}+{cw\over 2\sqrt{3}}+\sigma
z\right)\right]\int d\bar y\, e^{i\bar yz}\int dt\,t^{-p-1}\exp\left[{1\over
2}i\xi\left(t+{\zeta^2\over t}\right)\right]\,,
\label{B4}
\ee
where
\be
\xi={w^2-4v^2\over 12}\,,\qquad
\zeta=2\sqrt{3}i{\arccosh\sqrt{1+e^{2a}\bar y^2}\over\sqrt{w^2-4v^2}}\,.
\ee
From Ref.\ \cite{GR} (Eq.\ 8.421 No.\ 8, p.\ 966) the integral in $dt$ can be
evaluated. Using the new integration variable $\chi=\arccosh\sqrt{1+e^{2a}\bar
y^2}$ Eq.\ (\ref{B4}) is cast in the form
\ba
\Psi(a,b,c,\sigma)&=&N\exp\left[i\left({bv\over\sqrt{3}}+{cw\over 2\sqrt{3}}+\sigma
z\right)\right]e^{-a}\cdot\\
&\cdot&\int d\chi\,\cosh\chi \chi^{-p}
H_p^{(1)}\left(i\chi{\sqrt{w^2-4v^2}\over 2\sqrt{3}}\right)
e^{ize^{-a}\sinh\chi}\,.
\label{B6}
\ea
Choosing $p=-1/2$ and recalling that $H_{-1/2}^{(1)}(z)=\sqrt{2/\pi z}e^{iz}$
(see Ref.\ \cite{GR} Eq.\ 8.469, No.\ 6, p.\ 978) we have
\be
\displaystyle
\Psi(a,b,c,\sigma)=N\exp\left[i\left({bv\over\sqrt{3}}+{cw\over 2\sqrt{3}}+\sigma
z\right)\right]e^{-a}\sum_{\pm}
\int {d\omega\over\sqrt{1+\omega^2}}e^{-\left({\sqrt{w^2-4v^2}\over
2\sqrt{3}}\pm 1\right)\arcsinh\omega+ize^{-a}\omega}\,,
\label{B7}
\ee 
where $\omega=\sinh\chi$. Finally, from Eq.\ 3.483, p.\ 387, of Ref.\ \cite{GR}
we find 
\ba
\Psi(a,b,c,\sigma)&=&N\exp\left[i\left({bv\over\sqrt{3}}+{cw\over 2\sqrt{3}}+\sigma
z\right)\right]\left(ze^{-a}K_{\nu+1}(ze^{-a})+ze^{-a}K_{\nu-1}(ze^{-a})\right)
\nonumber\\
&=&N\exp\left[i\left({bv\over\sqrt{3}}+{cw\over 2\sqrt{3}}+\sigma
z\right)\right]K_{\nu}(ze^{-a})\,,
\label{B8}
\ea 
where
\be
\nu={\sqrt{w^2-4v^2}\over 2\sqrt{3}}\,,
\ee
and we have used Eq.\ 8.471, p.\ 979, of Ref.\ \cite{GR}. Setting
$v=\pm\sqrt{3}k_b$, $w=\pm 2\sqrt{3}k_c$, and $z=\pm k_\sigma$
($k_b,k_c,k_\sigma>0$) we obtain Eq.\ (\ref{WDW-solII}).

\end{document}